\documentclass[reprint,superscriptaddress,amsmath,amssymb,aps,pra]{revtex4-1}

\usepackage{graphicx}
\usepackage{dcolumn}
\usepackage{bm}
\usepackage{hyperref}
\usepackage{commath}

\usepackage{color}
\usepackage{upgreek}
\newcommand{\av}[1]{\langle #1 \rangle}
\newcommand{\ct}{\cos \theta}
\newcommand{\st}{\sin \theta}
\newcommand{\cct}{\cos^2 \theta}
\newcommand{\sst}{\sin^2 \theta}
\newcommand{\ccct}{\cos^3 \theta}
\newcommand{\ssst}{\sin^3 \theta}
\newcommand{\cccct}{\cos^4 \theta}
\newcommand{\sssst}{\sin^4 \theta}
\newcommand{\ccccct}{\cos^5 \theta}
\newcommand{\ssssst}{\sin^5 \theta}
\newcommand{\muT}{\upmu {\rm T}}

\begin{document}

\preprint{APS/123-QED}

\title{Magnetic field uniformity in neutron electric dipole moment experiments}

\def\JEDI{JEDI institute for the Force}
\def\LPC{Normandie Univ, ENSICAEN, UNICAEN, CNRS/IN2P3, LPC Caen, Caen, France}
\def\LPSC{Univ. Grenoble Alpes, CNRS, Grenoble INP, LPSC-IN2P3, Grenoble, France}
\def\CSNSM{CSNSM, Universit\'e Paris Sud, CNRS/IN2P3, Universit\'e Paris Saclay, Orsay-Campus, France}
\def\PSI{Paul Scherrer Institute (PSI), Villigen, Switzerland}
\def\ETH{ETH Z\"urich, Institute for Particle Physics and Astrophysics, Z\"urich, Switzerland}
\def\BERN{Laboratory for High Energy Physics, Albert Einstein Center for Fundamental Physics, University of Bern, Bern, Switzerland}
\def\KULEUVEN{Instituut voor Kern-- en Stralingsfysica, Katholieke~Universiteit~Leuven, Leuven, Belgium}
\def\SUSSEX{Department of Physics and Astronomy, University of Sussex, Brighton, United Kingdom}
\def\UNIFR{University of Fribourg, Fribourg, Switzerland}
\def\JUC{Marian Smoluchowski Institute of Physics, Jagiellonian University, Cracow, Poland}
\def\HNI{Henryk Niewodniczanski Institute of Nuclear Physics, Polish Academy of Sciences, Cracow, Poland}
\def\KCGUM{Institute of Nuclear Chemistry, Johannes Gutenberg University Mainz, Mainz, Germany}
\def\PTB{Physikalisch Technische Bundesanstalt, Berlin, Germany}
\def\KENT{University of Kentucky, Lexington, USA}
\def\UNIFR{Physics Department, University of Fribourg, Fribourg, Switzerland}

\def\RAL{Rutherford Appleton Laboratory, Chilton, Didcot, Oxon, United Kingdom}
\def\ILL{Institut Laue-Langevin, Grenoble, France}


\author{C.~Abel}				\affiliation{\SUSSEX}
\author{N.~J.~Ayres}			\affiliation{\SUSSEX}
\author{T.~Baker} 				\affiliation{\RAL}
\author{G.~Ban}					\affiliation{\LPC}
\author{G.~Bison}				\affiliation{\PSI}
\author{K.~Bodek}				\affiliation{\JUC}
\author{V.~Bondar}				\affiliation{\KULEUVEN}
\author{C.~B.~Crawford}			\affiliation{\KENT}
\author{P.-J.~Chiu}				\affiliation{\PSI}
\author{E.~Chanel}				\affiliation{\BERN}
\author{Z.~Chowdhuri}			\affiliation{\PSI}
\author{M.~Daum}				\affiliation{\PSI}
\author{B.~Dechenaux}			\affiliation{\LPC}
\author{S.~Emmenegger}			\affiliation{\ETH}
\author{L.~Ferraris-Bouchez}	\affiliation{\LPSC}
\author{P.~Flaux}				\affiliation{\LPC}
\author{P.~Geltenbort} 			\affiliation{\ILL}
\author{K.~Green} 				\affiliation{\RAL}
\author{W.~C.~Griffith}			\affiliation{\SUSSEX}
\author{M.~van~der~Grinten} 	\affiliation{\RAL}
\author{P.G.~Harris}			\affiliation{\SUSSEX}
\author{R.~Henneck}				\affiliation{\PSI} 
\author{N.~Hild}				\affiliation{\PSI}
\author{P.~Iaydjiev} 			\affiliation{\RAL}
\author{S.~N.~Ivanov} 			\affiliation{\RAL}
\author{M.~Kasprzak}			\affiliation{\PSI}
\author{Y.~Kermaidic}			\affiliation{\LPSC}
\author{K.~Kirch}				\affiliation{\PSI}\affiliation{\ETH} 
\author{H.-C.~Koch}				\affiliation{\PSI}\affiliation{\ETH} 
\author{S.~Komposch}			\affiliation{\PSI}\affiliation{\ETH} 
\author{P.~A.~Koss}				\affiliation{\KULEUVEN}
\author{A.~Kozela}				\affiliation{\HNI}
\author{J.~Krempel}				\affiliation{\ETH}
\author{B.~Lauss}				\affiliation{\PSI}
\author{T.~Lefort}				\affiliation{\LPC}
\author{Y.~Lemiere}				\affiliation{\LPC}
\author{A.~Leredde}				\affiliation{\LPSC}
\author{P.~Mohanmurthy}			\affiliation{\PSI}\affiliation{\ETH}
\author{D.~Pais}				\affiliation{\PSI}
\author{F.~M.~Piegsa}			\affiliation{\BERN}
\author{G.~Pignol}				
\email[Corresponding author: ]{guillaume.pignol@lpsc.in2p3.fr}
\affiliation{\LPSC}
\author{G.~Qu\'em\'ener}		\affiliation{\LPC}
\author{M.~Rawlik}				\affiliation{\ETH}
\author{D.~Rebreyend}			\affiliation{\LPSC}
\author{D.~Ries}				\affiliation{\KCGUM}
\author{S.~Roccia}				\affiliation{\CSNSM}
\author{D.~Rozpedzik}			\affiliation{\JUC}
\author{P.~Schmidt-Wellenburg}	\affiliation{\PSI}
\author{A.~Schnabel}			\affiliation{\PTB}
\author{N.~Severijns}			\affiliation{\KULEUVEN}
\author{R.~Virot}				\affiliation{\LPSC}
\author{A.~Weis}				\affiliation{\UNIFR}
\author{E.~Wursten}				\affiliation{\KULEUVEN}
\author{G.~Wyszynski}			\affiliation{\JUC}
\affiliation{\ETH}
\author{J.~Zejma}				\affiliation{\JUC}
\author{G.~Zsigmond}			\affiliation{\PSI}


\date{\today}

\begin{abstract}
Magnetic field uniformity is of the utmost importance in experiments to measure the electric dipole moment of the neutron. 
A general parametrization of the magnetic field in terms of harmonic polynomial modes is proposed, going beyond the linear-gradients approximation. 
We review the main undesirable effects of nonuniformities: depolarization of ultracold neutrons, and Larmor frequency shifts of neutrons and mercury atoms. 
The theoretical predictions for these effects were verified by dedicated measurements with the single-chamber neutron electric-dipole-moment apparatus installed at the Paul Scherrer Institute.  
\end{abstract}

\pacs{Valid PACS appear here}%
\maketitle


\section{\label{sec:Intro}Introduction}
Discovering a nonzero electric dipole moment (EDM) of the neutron would have far-reaching implications.
Indeed, the existence of an EDM for a simple spin 1/2 particle implies the violation of time-reversal invariance and therefore the violation of CP symmetry.
So far, the observed T and CP violation in nature is entirely accounted for by the Kobayashi Maskawa mechanism.
This mechanism predicts an unmeasurably small value for the EDMs of all subatomic particles.
Therefore, electric dipole moments are sensitive probes of new physics beyond the Standard Model of particle physics.
In fact, new CP violating interactions are needed to explain the generation of the matter-antimatter asymmetry in the early Universe.
Thus, the motivation to search for the neutron EDM lies at the interface between particle physics and cosmology.
The subject is treated in the classic book \cite{Khriplovich1997}. 
The connections between fundamental neutron physics and cosmology are treated in \cite{Abele2008,Dubbers2011,Pignol2015}. 
See also \cite{Pospelov2005,Lamoreaux2009,Engel2013,Jungmann2013,Schmidt-Wellenburg2016,Yamanaka2017,Filippone2018,Chupp2019} for recent reviews on EDMs. 

Since the first experiment by Smith, Purcell and Ramsey in 1951 \cite{Smith1957},
the precision on the neutron EDM has been improved by six orders of magnitude,
and yet the most recent measurement \cite{Pendlebury2015} is still compatible with zero:
\begin{equation}
d_n = ( -0.21 \pm 1.82 ) \times 10^{-26} \, e \, {\rm cm}.
\end{equation}

This result was obtained with an apparatus operated at the Institut Laue Langevin (ILL) built by the Sussex/RAL/ILL collaboration \cite{Baker2014}. 
As with almost all other contemporary or future nEDM projects, this experiment used ultracold neutrons (UCNs) stored for several minutes in a material bottle.
The bottle, a cylindrical chamber of height $12$~cm and diameter $47$~cm, sits in a stable and uniform vertical magnetic field with a magnitude of $B_0 = 1~\muT$.
In addition, a strong ($E \approx 10$~kV/cm) electric field is applied, either parallel or anti-parallel to the magnetic field.
One precisely measures the Larmor precession frequency $f_n$ of neutron spins in the chamber with Ramsey's method of separated oscillatory fields.
By comparing the neutron precession frequency in parallel and anti-parallel fields, one extracts
$d_n = \pi \hbar (f_{n, \uparrow \downarrow} - f_{n, \uparrow \uparrow}) / 2E$.

In these experiments, besides maximizing the number of stored ultracold neutrons,
the control of the magnetic field is the most important experimental challenge.
The time fluctuations of the magnetic field must be minimized and monitored, and the magnetic field should be sufficiently uniform. 
Even if external perturbations of the magnetic field are attenuated by several layers of shielding,
residual time variations of the $B_0$ field still need to be monitored in real-time. 
To this aim, the experiment \cite{Pendlebury2015, Baker2014} uses a co-magnetometer:
Spin-polarized $^{199}$Hg atoms fill the chamber, co-located with the stored ultracold neutrons \cite{Green1998,Ban2018}. 
The time-averaged precession frequency of the mercury spins $f_{\rm Hg}$ over each measurement cycle is used to correct for the drifts of the magnetic field through the relation $f_{\rm Hg} = \gamma_{\rm Hg}/(2\pi) B_0$, where
$\gamma_{\rm Hg}$ is the gyromagnetic ratio. 

Not only must the field be stable, with its time variations precisely monitored, it also needs to be extremely uniform over a large volume. 
As will be explained later, a field uniformity at a level better than $1$~nT must be achieved inside the chamber. 
For the purpose of tuning and characterizing the field uniformity, the co-magnetometer alone is not sufficient. 
One must therefore rely upon offline mapping of the magnetic field in the inner part of the apparatus, and/or upon an array of magnetometers around the chamber measuring the field in real time. 

In this paper, we discuss the effects of magnetic field nonuniformities in experiments measuring the neutron EDM with stored ultracold neutrons.
Specific concerns associated with the use of an atomic co-magnetometer are also dealt with in detail. In particular, 
the formalism described in the article is adequate to discuss the systematic effects in the experiment that was in operation at the Paul Scherrer Institute (PSI) during the period 2009-2017. 
The apparatus was an upgraded version of the one previously installed at the ILL that produced the current lowest experimental limit. 
However, we aim at a general treatment of the subject - whenever possible - so that the results are of interest for other past experiments such as \cite{Serebrov2015} as well as for the future experiments currently in development at the US Spallation Neutron Source \cite{Ito2007}, FRMII/ILL \cite{Altarev2012}, TRIUMF \cite{Picker2017}, PNPI \cite{Serebrov2017}, LANL and PSI \cite{Abel2018}. 

In the first part we present a general parameterization of the field in terms of a polynomial expansion.
It goes beyond the usual description in terms of linear gradients, a refinement that becomes necessary to quantify the systematic effects at the current level of sensitivity.
In the second and third parts, we discuss the effects of field nonuniformities on the statistical and systematic precision, respectively. 
Dedicated measurements were performed to corroborate the theoretical predictions for these effects. 

This article has two companion papers and should be read as the first episode of a trilogy. 
The second episode will describe the array of atomic Cesium magnetometers developed for the PSI nEDM experiment and the methods to optimize \emph{in situ} the field uniformity. 
The third episode will present the offline characterization of the magnetic field uniformity in the apparatus with an automated field-mapping device. 

\section{Harmonic polynomial expansion of the magnetic field}

In modern nEDM experiments a weak magnetic field $B_0 ~\approx 1~\muT$ is applied in a volume of about a cubic meter or more.
In the context of this article the field can be considered to be purely static.
The field $\vec{B}(x,y,z) \approx B_0 \vec{e}_z$ is very uniform, but the remaining nonuniformities have paramount consequences.
An adequate description of the nonuniformities is needed to discuss these consequences.

We construct a polynomial expansion (in terms of the Cartesian coordinates $x,y,z$)
of the magnetic field components, in the form
\begin{equation}
\label{HarmonicPolynomialExpansion}
\vec{B}(\vec{r}) = \sum_{l,m} G_{l,m}
\left(
\begin{array}{c}
\Pi_{x,l,m}(\vec{r}) \\
\Pi_{y,l,m}(\vec{r}) \\
\Pi_{z,l,m}(\vec{r}) \end{array}
\right)
\end{equation}
where the functions (or \emph{modes}) $\vec{\Pi}_{l,m}$ are harmonic polynomials in $x,y,z$ of degree $l$ and $G_{l,m}$ are the expansion coefficients. 

\begin{table}
\caption{
Associated Legendre polynomials up to $l = 5$. 
\label{associatedLegendre}}
\begin{tabular}{lll}
$l$ & $m$ & $P_l^m(\ct)$ \\
\hline \hline
1  & 0  & $\ct$ \\
1  & 1  & $-\st$ \\
\hline
2  & 0  & $\frac{1}{2}(3\cct-1)$ \\
2  & 1  & $-3 \ct \st$ \\
2  & 2  & $3 \sst$ \\
\hline
3  & 0  & $\frac{1}{2} \ct (5 \cct - 3)$ \\
3  & 1  & $-\frac{3}{2}(5 \cct - 1)\st$ \\
3  & 2  & $15 \ct \sst$ \\
3  & 3  & $-15 \ssst$ \\
\hline
4  & 0  & $\frac{1}{8}(35 \cccct - 30 \cct + 3)$ \\
4  & 1  & $-\frac{5}{2} \ct (7 \cct-3) \st$ \\
4  & 2  & $\frac{15}{2} (7 \cct - 1) \sst$ \\
4  & 3  & $-105 \ct \ssst$ \\
4  & 4  & $105 \sssst$ \\
\hline
5  & 0  & $\frac{1}{8}(63 \ccccct - 70 \ccct + 15 \ct)$ \\
5  & 1  & $-\frac{15}{8}(21 \cccct - 14 \cct + 1) \st$ \\
5  & 2  & $\frac{105}{2} (3 \ccct - \ct) \sst$ \\
5  & 3  & $-\frac{105}{2} (9\cct -1) \ssst$ \\
5  & 4  & $945 \ct \sssst$ \\
5  & 5  & $-945 \ssssst$ \\
\hline
\end{tabular}
\end{table}

The polynomials however cannot be chosen arbitrarily, since the magnetic field must satisfy Maxwell's equations: 
$\vec{\nabla} \cdot \vec{B} = 0$ and $\vec{\nabla} \times \vec{B} = 0$, in a region with neither currents nor magnetization. 
This requirement is equivalent to enforcing that the field is the gradient of a magnetic potential,
$\vec{B}(\vec{r}) = \vec{\nabla} \Sigma (\vec{r})$, with the potential satisfying Laplace's equation $\Delta \Sigma = 0$.
Solutions of Laplace's equation are called harmonic functions.
Therefore, all possible polynomial field components of degree $l-1$
are exactly obtained by taking the gradient of all possible harmonic polynomials of degree $l$.
The so-called solid harmonics, expressed in spherical coordinates as
\begin{equation}
r^l Y_{l,m}(\theta, \phi) = \sqrt{\frac{2l+1}{4\pi}\frac{(l-m)!}{(l+m)!}} \ r^l P_l^m(\ct) e^{i m \phi},
\end{equation}
form a basis of complex homogeneous polynomials, with $l$ the degree of the polynomial and $m$ an integer in the range $-l \leq m \leq l$.
In this formula $Y_{l,m}$ are the standard spherical harmonics and $P_l^m$ are the associated Legendre polynomials (listed in table \ref{associatedLegendre}).

To construct our basis, we need to take the real and imaginary parts of the complex polynomials.
In addition, we use a different and convenient normalization of the polynomials and define
\begin{equation}
\label{magneticPotential}
\Sigma_{l,m} = C_{l,m}(\phi) r^l P_l^{|m|}(\ct),
\end{equation}
with
\begin{eqnarray}
C_{l,m}(\phi) & = & \frac{(l-1)! (-2)^{|m|}}{(l+|m|)!} \cos(m \phi) \quad {\rm for } \quad m \geq 0 \\
\nonumber
C_{l,m}(\phi) & = & \frac{(l-1)! (-2)^{|m|}}{(l+|m|)!} \sin(|m| \phi) \quad {\rm for } \quad m < 0.
\end{eqnarray}
Finally, the modes are obtained by calculating the gradient of the magnetic potential:
\begin{equation}
\Pi_{x, l, m} = \partial_x \Sigma_{l+1,m} ; \ \Pi_{y, l, m} = \partial_y \Sigma_{l+1,m} ; \ \Pi_{z, l, m} = \partial_z \Sigma_{l+1,m}.
\end{equation}
Note that $l$ always refers to the degree of the polynomial, and therefore $\vec{\Pi}_{l, m}$ is obtained from the magnetic potential $\Sigma_{l+1,m}$ with $l$ differing by one unit.

An explicit calculation of the first order modes in Cartesian coordinates, up to $l=3$, is shown in table \ref{adequate}. 
For the expression of the modes in cylindrical coordinates, see table  \ref{adequateCylindrical} in appendix \ref{appendixCylindrical}. 
A similar parameterization has been proposed in the context of the SNS nEDM project \cite{Nouri2014,Nouri2015}. 
See also ref. \cite{Maldonado2017} for the application of the scalar magnetic potential method in other precision experiments with polarized neutrons. 
In fact the use of spherical harmonics to describe a near-uniform field appeared first in the context of Nuclear Magnetic Resonance \cite{Anderson1961} and then in Magnetic Resonance Imaging \cite{Hoult1981,Briguet1985}, where field uniformity is also of great importance. 

When dealing with a perfectly uniform magnetic field, that field is described by the $l=0$ terms only and we simply have
\begin{eqnarray}
G_{0, -1} & = & B_y, \\
G_{0, 0} & = & B_z, \\
G_{0, 1} & = & B_x.
\end{eqnarray}
In the case of a field with uniform gradients, that field is described by the $l=0$ and $l=1$ terms and we have
\begin{eqnarray}
G_{1, -2} & = & \partial_y B_x = \partial_x B_y, \\
G_{1, -1} & = & \partial_y B_z = \partial_z B_y, \\
G_{1, 0} & = & \partial_z B_z = -\partial_x B_x - \partial_y B_y, \\
G_{1, 1} & = & \partial_x B_z = \partial_z B_x, \\
G_{1, 2} & = & \frac{1}{2} (\partial_x B_x-\partial_y B_y). 
\end{eqnarray}
The harmonic polynomial expansion of the field nonuniformities given by Eq.\ \eqref{HarmonicPolynomialExpansion} is a natural generalization of the description in terms of uniform gradients. 
The coefficients $G_{l,m}$ are the generalized gradients for the modes of degree $l$. 
Given the degree of maturity of nEDM experiments, this generalization is necessary to discuss the phenomena associated with field nonuniformity at the appropriate level of accuracy.

\begin{table*}
\caption{
The basis of harmonic polynomials sorted by degree.
\label{adequate}}
\begin{tabular}{cc|ccc}
$l$ & $m$ & $\Pi_x$ & $\Pi_y$  & $\Pi_z$ \\
\hline \hline
$0$ & $-1$  & $0$                & $1$                & $0$            \\
$0$ & $0$  & $0$                & $0$                & $1$            \\
$0$ & $1$  & $1$                & $0$                & $0$            \\
\hline
$1$ & $-2$  & $y$                & $x$                & $0$            \\
$1$ & $-1$  & $0$                & $z$                & $y$            \\
$1$ & $0$  & $-\frac12 x$       & $-\frac12 y$       & $z$            \\
$1$ & $1$  & $z$                & $0$                & $x$            \\
$1$ & $2$  & $x$                & $-y$               & $0$            \\
\hline
$2$ & $-3$  & $2xy$              & $x^2-y^2$          & $0$            \\
$2$ & $-2$  & $2yz$              & $2xz$              & $2xy$          \\
$2$ & $-1$  & $-\frac12 xy$      & $-\frac14 (x^2+3y^2-4z^2)$    & $2yz$   \\
$2$ & $0$  & $-xz$              & $-yz$              & $z^2 - \frac12 (x^2+y^2)$ \\
$2$ & $1$  & $-\frac14 (3x^2+y^2-4z^2)$  & $-\frac12 xy$ & $2xz$ \\
$2$ & $2$  & $2xz$              & $-2yz$             & $x^2-y^2$      \\
$2$ & $3$  & $x^2-y^2$          & $-2xy$             & $0$      \\
\hline
$3$ & $-4$ & $3x^2y-y^3$         & $x^3-3xy^2$        & $0$  \\
$3$ & $-3$ & $6xyz$              & $3(x^2z-y^2z)$     & $3x^2y-y^3$  \\
$3$ & $-2$ & $-\frac12 (3x^2y+y^3-6yz^2)$  & $-\frac12 (x^3+3xy^2-6xz^2)$ & $6xyz$  \\
$3$ & $-1$ & $-\frac32 xyz$      & $-\frac14 (3x^2z+9y^2z-4z^3)$ & $3yz^2 - \frac34 (x^2y+y^3)$  \\
$3$ & $0$ & $\frac38 (x^3+xy^2-4xz^2)$  & $\frac38 (x^2y+y^3-4yz^2)$ & $z^3-\frac32 z (x^2+y^2)$  \\
$3$ & $1$ & $-\frac14 (9x^2z+3y^2z-4z^3)$  & $-\frac32 xyz$ & $3xz^2 - \frac34 (x^3+xy^2)$  \\
$3$ & $2$ & $-x^3+3xz^2$        & $-3yz^2+y^3$       & $3 (x^2z-y^2z)$  \\
$3$ & $3$ & $3(x^2z-y^2z)$      & $-6xyz$            & $x^3-3xy^2$  \\
$3$ & $4$ & $x^3-3 x y^2$       & $-3x^2y+y^3$       & $0$  \\
\hline
\end{tabular}
\end{table*}

\section{Field uniformity and statistical precision: neutron depolarization}

We now discuss the effects of a nonuniform magnetic field on the statistical uncertainty,
which is limited by the precision of the determination of the neutron precession frequency $f_n$. 
The measurement of $f_n$ uses Ramsey's method of separated oscillatory fields. 
In short, a chamber is first filled with polarized ultracold neutrons, and then a $\pi/2$ pulse is applied to the neutron spins using a transverse oscillating field. The neutron spins then precess in the transverse plane for a precession time $T$. 
Finally a second $\pi/2$ pulse is applied, and the chamber is then opened to count the number of spin-up and spin-down neutrons. 
The asymmetry in the counting depends on the difference between the applied frequency (used to generate the pulses) and the Larmor frequency $f_n$ (to be measured). 
With this method the statistical uncertainty on the neutron EDM, due to Poisson fluctuations of the neutron counts, is: 
\begin{equation}
\label{FOM}
\sigma d_n = \frac{\hbar}{2 \alpha E T \sqrt{N}},
\end{equation}
where $E$ is the electric-field strength, $N$ is the total number of neutrons measured during the measurement sequence and
$\alpha$ is the visibility - or contrast - of the Ramsey resonance, which refers to the polarization of the ultracold neutrons at the end of the precession period multiplied by the analyzing power of the spin analyzer system.
In order to keep the visibility $\alpha$ as high as possible, all the depolarization mechanisms at play during the precession time must be understood and minimized.
Typically, in the current experiment at PSI with a single chamber, we achieved $\alpha \approx 0.75$ after a precession time of $T = 180$~s.

In previous works \cite{Harris2014,Afach2015_PRL, Afach2015_PRD}, we have identified the main mechanisms responsible for the decay of the neutron polarization while they are stored in the chamber.
The variation of $\alpha$ with respect to the precession duration can be written as a sum of three contributions:
\begin{equation}
\label{transverseDepol}
\frac{d \alpha}{dT} = - \frac{\alpha}{T_{2, \rm wall}} - \frac{\alpha}{T_{2, \rm mag}} + \dot{\alpha}_{\rm grav},
\end{equation}
where $T_{2, \rm wall}$ is the transverse spin relaxation time due to wall collisions (see Sec. \ref{wall_depolarization}), 
$T_{2, \rm mag}$ is the transverse spin relaxation time due to intrinsic depolarization in a nonuniform field (see Sec. \ref{intrinsic_depolarization}),
and $\dot{\alpha}_{\rm grav}$ the contribution from gravitationally enhanced depolarization (see Sec. \ref{enhanced_depolarization}). 
Note that Eq.\ \eqref{transverseDepol} applies to spins that are precessing in the magnetic field; this process is called \emph{transverse depolarization}. 
The corresponding situation for when spins are aligned along the holding field is called \emph{longitudinal depolarization}. 
In this case the depolarization rate $1/T_1$ also receives contributions from wall collisions and field nonuniformities as
\begin{equation}
\label{longitudinalDepol}
\frac{1}{T_1} = \frac{1}{T_{1, \rm wall}} + \frac{1}{T_{1, \rm mag}}, 
\end{equation}
and it is in general different from the transverse depolarization rate. 
We will now review all of these mechanisms in more detail.

\subsection{Wall depolarization}
\label{wall_depolarization}

When colliding with the wall of the precession chamber, a neutron can have its spin affected by magnetic impurities contained within the wall.
Given that the interaction time with the wall is much shorter than the Larmor precession period, and that any orientation of the spin is equally affected on average, we can anticipate that the transverse and longitudinal relaxation rates will be identical:
\begin{equation}
\frac{1}{T_{2, \rm wall}} = \frac{1}{T_{1, \rm wall}} = \beta \nu, 
\end{equation}
where $\beta$ is the depolarization probability per wall collision and $\nu$ is the average frequency of wall collisions. 
Suitable materials have depolarization probabilities in the range $10^{-6} \lesssim \beta \lesssim 10^{-5}$ 
(see \cite{Bondar2017} for a recent work on wall depolarization). 
In practice the wall collision frequency is less than $50$~s$^{-1}$, and  
$T_1$ is generally measured to be longer than 2000~s. 
Therefore, although wall depolarization is not a negligible process, it does not constitute a serious limitation for maintaining a high polarization.

\subsection{Gravitationally enhanced depolarization}
\label{enhanced_depolarization}

Ultracold neutrons are neutrons of extremely low kinetic energy, typically 200~neV or less.
They are therefore significantly affected by gravity: different energy groups of neutrons have different mean heights in the chamber.
In the presence of a vertical field gradient, the spins of neutrons in different energy groups precess at a slightly different rate, resulting in a phenomenon referred to as \emph{gravitationally enhanced depolarization}. 
This mechanism concerns the transverse depolarization only. 

For a quantitative description of the effect, we assume that the field can be described by the polynomial expansion up to order $l=1$.
We denote the probability for a neutron to belong to the energy group $\epsilon$ as $n(\epsilon) d \epsilon$.
After the precession time $T$, spins belonging to the energy group $\epsilon$ have accumulated a phase difference, with respect to the average phase of  all neutrons, of
\begin{equation}
\varphi(\epsilon,T) = \gamma_n G_{1,0} (\bar{z}(\epsilon) - \langle z \rangle) T, 
\end{equation}
where $\bar{z}(\epsilon)$ is the mean height of neutrons in this group, $\langle z \rangle$ is the mean height of the whole ensemble of neutrons and $\gamma_n$ is the neutron gyromagnetic ratio. 
Assuming that each group of neutrons is initially perfectly polarized, and neglecting the depolarization within a group, the final polarization after the precession time $T$ is
\begin{equation}
\alpha(T) = \int \cos \varphi(\epsilon,T) n(\epsilon) d\epsilon. 
\end{equation}
For small values of the phase (which is generally the case for small gradients) the cosine can be approximated using a second-order Taylor expansion:
\begin{equation}
\alpha(T) = 1 - \frac{1}{2}\int \varphi(\epsilon,T)^2 n(\epsilon) d\epsilon.
\end{equation}
Finally, the depolarization rate $\dot{\alpha}_{\rm grav}$ is obtained from the derivative of the previous expression over precession time: 
\begin{equation}
\label{GravDepol}
\dot{\alpha}_{\rm grav} = - \gamma_n^2 G_{1,0}^2 {\rm Var}[\bar{z}] \ T,
\end{equation}
with ${\rm Var}[\bar{z}]$ the variance of the distribution of $\bar{z}(\epsilon)$:
\begin{equation}
{\rm Var}[\bar{z}] = \int (\bar{z}(\epsilon) - \langle z \rangle)^2 n(\epsilon) d\epsilon.
\end{equation}

\subsection{Intrinsic depolarization}
\label{intrinsic_depolarization}

The intrinsic depolarization refers to the decay of polarization within an energy group.
It is due to the fact that different neutrons in a group have different random trajectories in a nonuniform field and therefore different histories of the magnetic field $\vec{B}(t)$. 
This process can be described by spin-relaxation theory, which is a general approach to calculate frequency shifts and relaxation rates on a quantum system in terms of the correlation function of the disturbance, to second order in the disturbance. 
In our case the disturbances are the field components $B_i(t)$ with $i \in \{x,y,z \}$, and their correlation functions $\langle B_i(t_1) B_j(t_2) \rangle$ are the ensemble averages of the quantities $B_i(t_1) B_j(t_2)$ over the neutrons stored in the chamber. Here we assume that the motion of the neutrons in the chamber is stationary in the statistical sense and therefore $\langle B_i(t_1) B_j(t_2) \rangle = \langle B_i(0) B_j(t_2-t_1) \rangle$. 
Specifically, it is the deviation from the mean value of the field components, $B_i^c(t) = B_i(t) - \langle B_i \rangle$, that induces the relaxation of the spin. 
In the language of random processes, $B_i^c(t)$ is the \emph{centered} variable associated with $B_i(t)$; hence the notation with the exponent \emph{c}.

Applying the spin-relaxation theory to our problem of spin-1/2 particles in a bottle \cite{McGregor1990,Golub2011,Pignol2015b} , one finds
\begin{equation}
\frac{1}{T_{1, \rm mag}} = \gamma_n^2 \int_0^\infty \langle B_x^c(0) B_x^c(t) + B_y^c(0) B_y^c(t) \rangle \cos \omega t \, dt
\end{equation}
for the longitudinal relaxation rate and
\begin{equation}
\label{T2intrinsic1}
\frac{1}{T_{2, \rm mag}} = \frac{1}{2 T_{1, \rm mag}} + \gamma_n^2 \int_0^\infty \langle B_z^c(0) B_z^c(t)\rangle dt
\end{equation}
for the transverse relaxation rate.
In these expressions, $\omega = \gamma_n B_0$ is the angular Larmor precession frequency, and $\langle X \rangle$ refers to the ensemble average of the quantity $X$ over the neutrons stored in the chamber.

In fact, the depolarization induced by the field components $B_x$ and $B_y$ transverse to the holding field $B_0$ are very small.  
In the regime where the precession frequency $f_n$ 
is much higher than the wall collision frequency $\nu$, it has been shown in \cite{Afach2015_PRD} that the order of magnitude of the longitudinal depolarization rate can be estimated by
\begin{equation}
\frac{1}{T_{1, \rm mag}} \sim \frac{v^3 \Delta B_{\rm T}^2}{80 R^3 \gamma_n^2 B_0^4},
\end{equation}
where $v$ is the neutron speed, $R$ is the radius of the chamber (assumed to be cylindrical, with the axis aligned along $z$), and $\Delta B_{\rm T}$ is the typical value for the transverse field difference in the chamber. 
Note that a uniform transverse field has no effect. 
Using realistic numbers for the nEDM apparatus installed at PSI ($2R = 47$~cm, $B_0 = 1 \ \muT$, $v = 3$~m/s and $\Delta B_T = 2$~nT) we find $T_{1, \rm mag} \sim 10^{10}$~s. 
Therefore we will not give a precise description of the transverse depolarization in the harmonic polynomial expansion formalism. 

To calculate the intrinsic depolarization rate, it is justified to neglect transverse fields and keep only the effect of longitudinal nonuniformities.  
Expressing the field in the basis of harmonic polynomials, the correlation function becomes
\begin{equation}
\langle B_z^c(0) B_z^c(t)\rangle = \sum_{l,l',m, m'} G_{l,m}G_{l',m'} \langle \Pi_{z,l,m}^c(0) \Pi_{z,l',m'}^c(t)\rangle. 
\end{equation}
In the case of a cylindrical chamber, the terms with $m \neq m'$ cancel due to rotational symmetry around the cylinder axis. 
The intrinsic depolarization rate can then be expressed as 
\begin{equation}
\label{T2intrinsic2}
\frac{1}{T_{2, \rm mag}} = \gamma_n^2 \sum_{l,l',m} G_{l,m}G_{l',m} \int_0^\infty \langle \Pi_{z,l,m}^c(0) \Pi_{z,l',m}^c(t)\rangle dt. 
\end{equation}
At this point we can recognize that the depolarization rate is a quadratic function of the generalized gradients $G_{l,m}$, and that it depends on how fast a correlation of the type $\langle \Pi_{z,l,m}(0) \Pi_{z,l',m}(t)\rangle $ decays. 
In particular, slower neutrons depolarize more quickly. 
Also, for experiments using a mercury co-magnetometer, the mercury atoms depolarize in this fashion with a much slower rate than the neutrons because the mercury atoms are much faster. 

Now, for a precise calculation of the depolarization rate of ultracold neutrons in a given magnetic field gradient a Monte Carlo simulation of the trajectories of the neutrons can be used. 
Such a study, in the case $l=1$, has been presented in \cite{Afach2015_PRD}, together with an intuitive model of the depolarization in linear gradients. 
The intuitive model predicts 
\begin{equation}
\label{T2intrinsic3}
\frac{1}{T_{2, \rm mag}} = \frac{8R^3 \gamma_n^2}{9\pi v} (G_{1, -1}^2 + G_{1, 1}^2) + \frac{\mathcal{H}^3 \gamma_n^2}{16 v} G_{1, 0}^2, 
\end{equation}
where $v$ is the speed of the neutrons, $R$ is the radius of the storage chamber, and $\mathcal{H}$ is the maximum height of the neutrons of speed $v$. 
The intuitive model reproduces the Monte Carlo results quite well. 

\subsection{Experimental verification of the depolarization theory}

We have conducted dedicated measurements on gradient-induced neutron depolarization with the nEDM apparatus installed at the PSI ultracold neutron source \cite{Lauss2014, Bison2017}. 
In a first series of measurements, performed in May 2016, we varied in a controlled way the vertical gradient $G_{1,0}$ and measured the final neutron polarization after a storage time of $T = 180$~s. 
In a second series, performed in September 2017, we measured the final polarization as a function of the horizontal gradient $G_{1,1}$. 

At each cycle the precession chamber is filled with polarized neutrons. 
The neutrons are polarized by a 5~T superconducting magnet installed between the UCN source and the nEDM apparatus. 
Only one spin component is transmitted through the bore of the magnet, thereby polarizing the neutrons with an efficiency close to 100 \%. 
Three types of runs were recorded to measure the final polarization, corresponding to three types of storage conditions: 
\begin{enumerate}
\item Longitudinal polarization: neutrons are stored with their spin aligned with the holding magnetic field, and no spin-flip pulse is applied. 
During storage the polarization decreases at a rate given by Eq.\ \eqref{longitudinalDepol}. 
\item Ramsey: a $\pi/2$ pulse is applied at the beginning and at the end of the precession period (with a duration of 2~s each), so that the neutron spins precess in the holding field during the storage period. This is the normal mode of operation during nEDM runs because it allows a precise determination of the precession frequency. During precession the polarization decreases at the rate given by Eq.\ \eqref{transverseDepol}. 
\item Spin-echo: in addition to the two $\pi/2$ pulses applied at the beginning and at the end of the precession period, a $\pi$ pulse is applied exactly halfway through the precession time. 
The effect of the $\pi$ pulse is to cancel the dephasing of different neutron energy groups \cite{Afach2015_PRL}, and therefore the depolarization rate is given by $d\alpha/dT = -\alpha/T_{2, \rm wall} - \alpha/T_{2, \rm mag}$. 
This mode allows one to isolate the intrinsic transverse depolarization from the gravitationally enhanced depolarization. 
\end{enumerate}

At the end of the storage period the ultracold neutrons are released from the precession chamber by opening the UCN shutter, allowing them to proceed to the spin analyzer\cite{Afach2015_EPJA}. 
This device simultaneously counts the neutrons in each of the two spin states: it has two arms, each of which includes (i) an adiabatic spin flipper, (ii) a magnetized iron foil that transmits one spin component and reflects the other, (iii) a set of $^6$Li-doped glass scintillators \cite{Ban2016} to count the neutrons. 
Finally, the asymmetry 
\begin{equation}
A = \frac{N_{\uparrow}-N_{\downarrow}}{N_{\uparrow}+N_{\downarrow}}
\end{equation}
is calculated. 
The efficiency of the spin analyzer is not perfect due to the finite efficiency (about 90\%) of the magnetized foils. 

\begin{figure}
\includegraphics[width = \columnwidth]{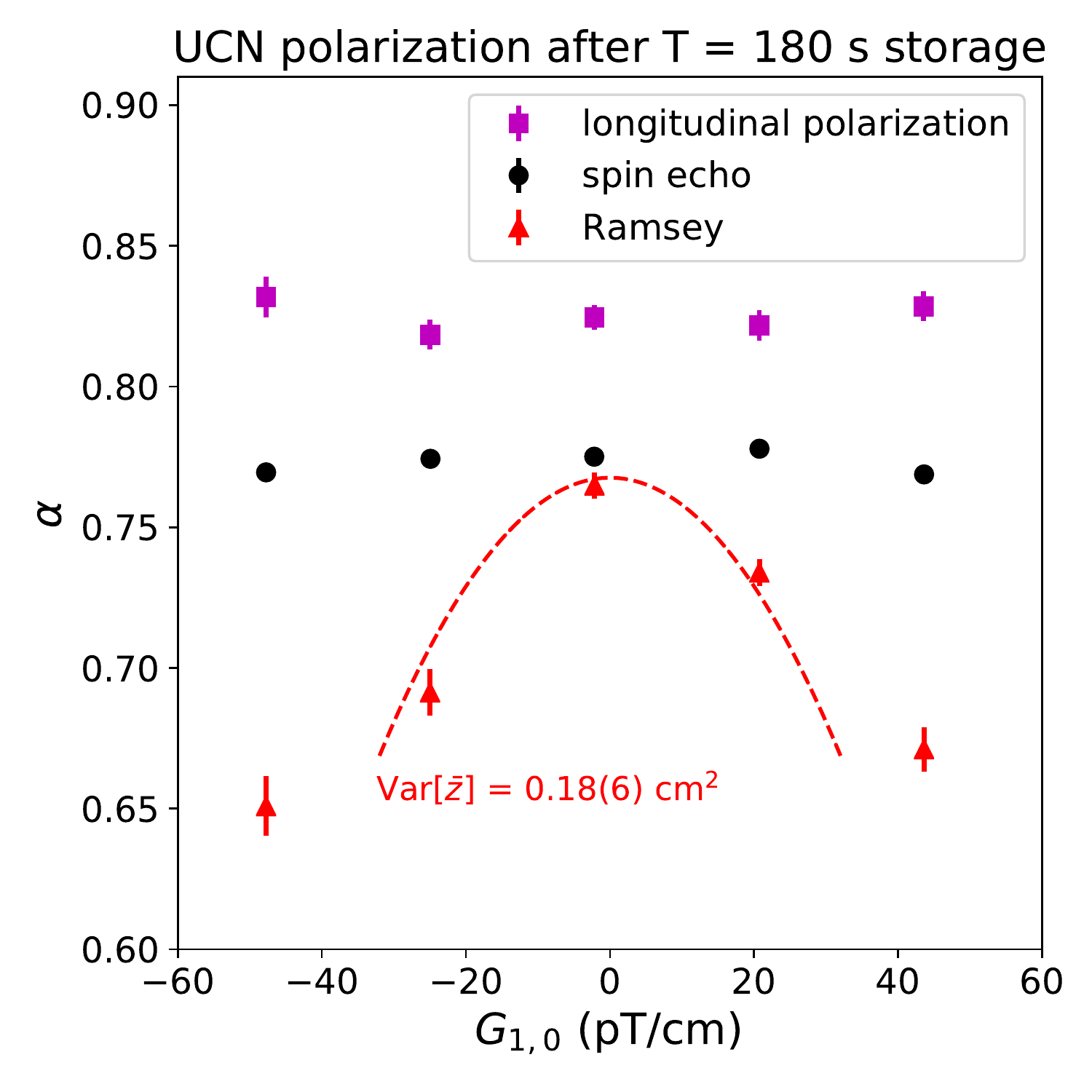}
\caption{\label{fig:G10scan}
Final polarization of ultracold neutrons after a storage time of $180$~s as a function of an applied vertical gradient $G_{1,0}$. 
Squares: longitudinal polarization; filled circles: polarization after a spin-echo run; triangles: polarization after a normal Ramsey run. 
The dashed line is a fit of the gravitationally enhanced depolarization model based on Eq.\ \eqref{GravDepol} to the data (excluding the two points at large gradients for which the small phase approximation is not valid). 
}
\end{figure}

For measurements in the longitudinal and spin-echo modes, the polarization is directly given by the asymmetry, i.e.\ $\alpha = A$. 
In the Ramsey mode, the polarization is given by the asymmetry at the resonance, i.e. $\alpha = A(f_{\rm RF} = f_n)$. 
In practice one measures the asymmetry as a function of the applied frequency $f_{\rm RF}$ of the $\pi/2$ pulses for several (typically eight) cycles and then fits the Ramsey fringe by a cosine function. 
The polarization $\alpha$ is given by the maximum - or visibility - of the Ramsey curve $A$ versus $f_{\rm RF}$. 

The gradients $G_{1,0}$ or $G_{1,1}$ are applied by setting well-defined currents in the set of correcting coils. 
The gradients are measured in real time with an array of cesium magnetometers. 

Figure \ref{fig:G10scan} shows the results of a measurement of 
the final polarization as a function of an applied vertical gradient $G_{1,0}$. 
Within the range of applied gradients, $|G_{1,0}| < 50$~pT/cm, the longitudinal polarization and the spin-echo polarization are constant. 
This is consistent with the expectation from Eq.\ \ref{T2intrinsic3} that the intrinsic magnetic depolarization is too small to be measured. 
The fact that the spin-echo polarization is smaller than the longitudinal polarization could be explained by possible residual horizontal gradients of the type $G_{1,1}$. 
We observe gravitationally enhanced depolarization in the Ramsey mode, with the polarization decreasing under the application of a finite gradient. 
We fit the model 
$\alpha(G_{1,0}) = \alpha_0 - \frac12 \gamma_n^2 G_{1,0}^2 {\rm Var}[\bar{z}] T^2$
to the data with $\alpha_0$ and ${\rm Var}[\bar{z}]$ as free parameters. 
We find ${\rm Var}[\bar{z}] = 0.18 \pm 0.06$~cm$^2$, a plausible value for stored ultracold neutrons. 

\begin{figure}
\includegraphics[width = \columnwidth]{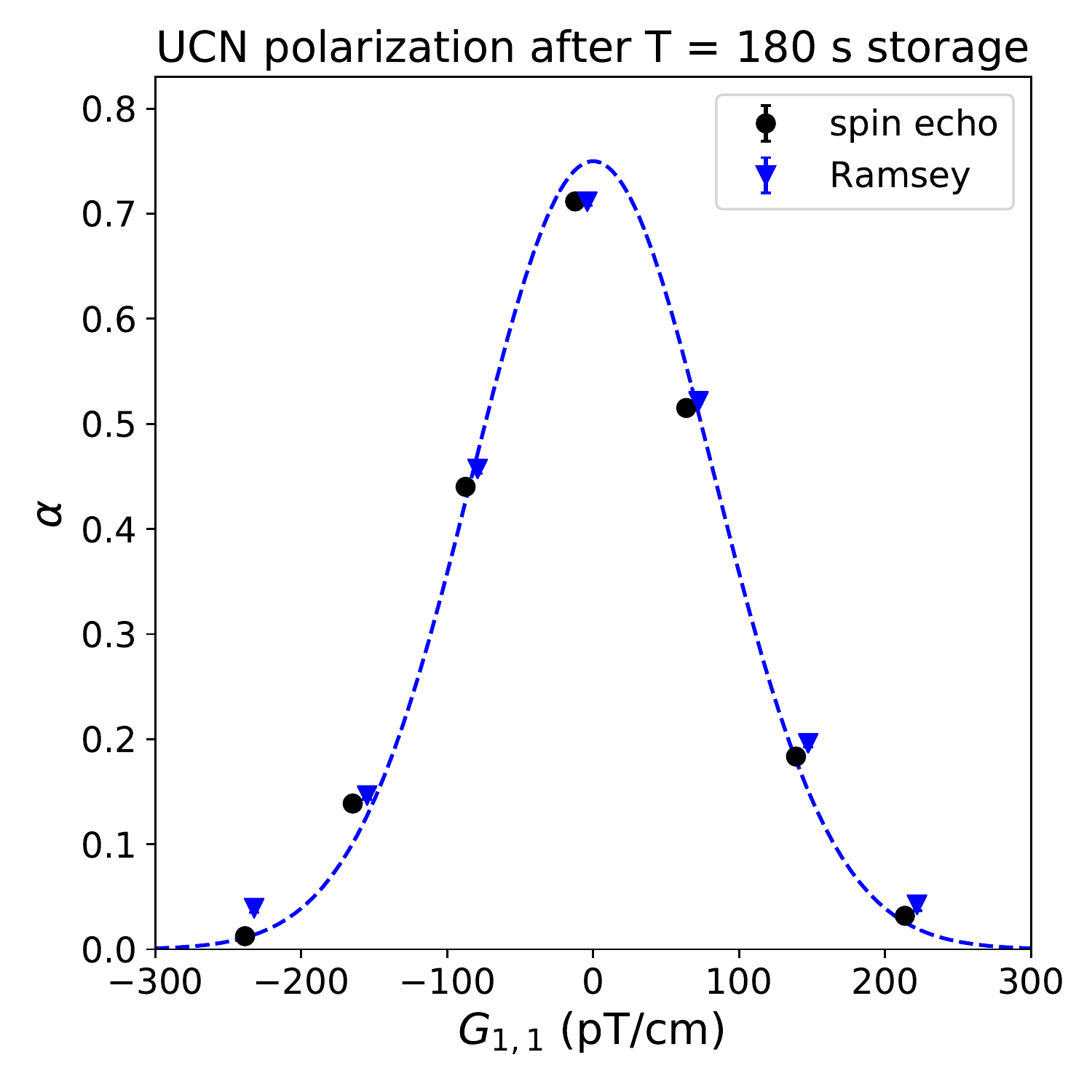}
\caption{\label{fig:G11scan}
Final polarization of ultracold neutrons after a storage time of $180$~s as a function of an applied horizontal gradient $G_{1,1}$. 
Filled circles: polarization after a spin-echo run; triangles: polarization after a normal Ramsey run. 
The dashed line correspond to the model Eq.\ \eqref{T2intrinsic4} with $\alpha_0 = 0.75$ and $v = 3$~m/s. 
}
\end{figure}

Figure \ref{fig:G11scan} shows the result of scanning the horizontal gradient $G_{1,1}$. The precession time was kept constant at $T = 180$~s. 
In this case, as expected, the applied gradient affects the polarization in the same manner as for the spin-echo and Ramsey runs. 
We have plotted (dashed line) the expected dependence
\begin{equation}
\label{T2intrinsic4}
\alpha(G_{1,1}) = \alpha_0 \exp \left( - \frac{T}{T_{2, \rm mag}(G_{1,1})} \right), 
\end{equation}
where $T_{2, \rm mag}(G_{1,1})$ is given by the intuitive model Eq.\ \eqref{T2intrinsic3} and we have chosen the parameters $\alpha_0 = 0.75$ and $v = 3$~m/s. 

Clearly, the data from the $G_{1,0}$ and $G_{1,1}$ scans are in good qualitative agreement with the expectations. 
There are two different mechanisms at play. 
The horizontal gradient $G_{1,1}$ induces a truly irreversible depolarization process, since the polarization cannot be recovered by the spin-echo method. 
On the other hand, the vertical gradient $G_{1,0}$ mainly affects the polarization through a loss of coherence of different energy groups separated by gravity; this coherence can be recovered through the spin-echo technique. 

\section{Field uniformity and systematic effects: frequency shifts}

In the present section we will cover the case of Larmor frequency shifts of particles --~ultracold neutrons or atoms~-- evolving in a nonuniform magnetic field in conjunction with an electric field.
We first review the linear-in-electric-field frequency shift, which constitutes an important direct systematic effect. 
In particular we calculate the false mercury EDM in terms of the coefficients of the harmonic expansion, and we discuss the effects of higher order modes. 
We will then review the electric-field-independent frequency shifts. 

\subsection{Motional false EDM}

When a particle moves with a velocity ${\bf v}$ through a static electric field ${\bf E}$, it experiences a (relativistic) motional magnetic field ${\bf B}_{\rm m} = {\bf E} \times {\bf v}/c^2$.
For trapped particles the velocity averages to zero, and therefore one is naively led to conclude that the effect vanishes.
This is indeed the case if the magnetic field is perfectly uniform.
However, when the particle spins evolve in a nonuniform magnetic field the motional field ${\bf B}_{\rm m}$ does induce a  linear-in-electric-field frequency shift $\delta f$. 
This effect has been extensively studied theoretically  \cite{Pignol2015b,Pendlebury2004,Lamoreaux2005,Barabanov2006,Clayton2011,Pignol2012,Swank2012,Steyerl2014,Golub2015,Swank2016}. 
The associated false EDM can be calculated in the framework of spin relaxation theory: 
\begin{equation}
\label{MotionalFalseEDM_General}
d^{\rm false} = \frac{\hbar \gamma^2}{2c^2} \int_0^\infty \langle B_x(0) v_x(t) + B_y(0) v_y(t) \rangle  \cos \omega t \, dt.
\end{equation}

Now, the magnitude of this undesirable false EDM critically depends on whether the particles are moving quickly or slowly, in a sense that we shall define.
With a mean square velocity $v_{\rm RMS} = \sqrt{\langle v_x^2 \rangle}$, it typically takes a time $\tau_c = 2R/v_{\rm RMS}$ for a particle to diffuse from one side of the chamber to the other ($2R$ is the typical transverse size of the chamber; for example, its diameter in the case of a cylindrical chamber).
After this time a correlation function of the type $\langle B(0) v(\tau_c) \rangle$ will have decayed to a small value.
The adiabaticity parameter is defined as $\omega \tau_c$.
For ultracold neutrons one usually has $\omega \tau_c \gg 1$, which means that the Larmor frequency is much faster than the wall collision rate: this is the \emph{adiabatic regime} of slow particles in a high field.
On the other hand, for mercury atoms at room temperature in a $B_0 = 1~\muT$ field $\omega \tau_c < 1$: this is the \emph{nonadiabatic regime} 
of fast particles in a low field.
In the adiabatic regime, the linear-in-electric-field frequency shift can be interpreted as originating from a geometric phase, as first noticed in \cite{Commins1991}. 
In fact the motional false EDM was called the \emph{geometric phase effect} in earlier publications. 

The general expression for the motional false EDM given in Eq.\ \eqref{MotionalFalseEDM_General} takes simplified forms in the adiabatic and nonadiabatic approximations:
\begin{eqnarray}
d^{\rm false} & = - \frac{\hbar v_{\rm RMS}^2}{2 c^2 B_0^2} \langle \frac{\partial B_z}{\partial z} \rangle & \quad {\rm (adiabatic)}    \\
\label{HgFalseEDM1}
d^{\rm false} & = -\frac{\hbar \gamma^2}{2 c^2} \langle x B_x + y B_y \rangle                              & \quad {\rm (nonadiabatic)},
\end{eqnarray}
where the brackets now refer to the volume average over the precession chamber.
It should be emphasized that these expressions are valid for an arbitrary form of the magnetic nonuniformity.

In the simple case of a uniform gradient, i.e. $G_{1,0} \neq 0$ and all other $G_{l,m}$ modes set to zero, in a cylindrical chamber of diameter $2R = 47$~cm, these expressions can be simplified for the neutron (adiabatic case) and mercury (nonadiabatic case) false EDM\footnote{We note a numerical error in the corresponding expression of equation 5 in \cite{Afach2015}}:
\begin{eqnarray}
d^{\rm false}_n & = & - \frac{\hbar v_{\rm RMS}^2}{2 c^2 B_0^2} G_{1,0} \\
 & \approx & - \frac{G_{1,0}}{\rm 1 pT/cm} \times 1.46 \times 10^{-28} e \, {\rm cm},
\end{eqnarray}
\begin{eqnarray}
d^{\rm false}_{\rm Hg} & = & \frac{\hbar \gamma_{\rm Hg}^2}{8 c^2} R^2 G_{1,0} \\                          & \approx & 
\label{HgFalseEDM}
\frac{G_{1,0}}{\rm 1 pT/cm} \times 1.15 \times 10^{-27} e \, {\rm cm}, 
\end{eqnarray}
the neutron case being calculated with $v_{\rm RMS} = 2$~m/s and with $B_0 = 1~\muT$. Because the mercury co-magnetometer is used to correct the neutron frequency for the drifts of the magnetic field, the false EDM of the mercury atoms translates to a false neutron EDM with a magnitude of
\begin{eqnarray}
\label{HgFalseEDM2}
d^{\rm false}_{n \leftarrow {\rm Hg}} & = & \abs{\frac{\gamma_n}{\gamma_{\rm Hg}}} d^{\rm false}_{\rm Hg} \\
& \approx & \frac{G_{1,0}}{\rm 1 pT/cm} \times 4.42 \times 10^{-27} e \, {\rm cm} .
\end{eqnarray}
It should be noted that the mercury-induced false neutron EDM is much larger than the directly induced neutron motional false EDM. 

In fact, it can be shown that the false EDM of a trapped particle is maximum at zero magnetic field, i.e.\ in the nonadiabatic limit. 
This explains why the mercury co-magnetometer running at $B_0 = 1 \, \muT$ is a source of large systematic effects. 
It should be said that, despite the existence of such (by now well understood) effects, the use of a co-magnetometer for these measurements is truly invaluable, and in its absence the credibility of any results might well be brought into question. 
Some compensation can be achieved through use of a double chamber, with electric fields in opposite directions and each chamber effectively acting as a magnetometer for the other, but this still does not truly sample the co-located field in a precise way. 
For a large-scale cryogenic experiment, for example, an alternative that has been proposed to the room-temperature mercury co-magnetometer is the concept of a helium-3 co-magnetometer diluted in superfluid helium-4 bath, for which the false EDM can be set to zero by adjusting the temperature of the bath \cite{Barabanov2006}. 
At room temperature, though, another alternative that has recently been proposed by one of us is to operate the mercury co-magnetometer at a higher ``magic'' magnetic field to set the false EDM to zero \cite{Pignol2019}. 
While this is an attractive possibility for a future experiment, it brings with it significant difficulties in ensuring the uniformity of the magnetic field to the level required to avoid depolarization of the neutrons. 
In the remainder of the present paper we will consider the nonadiabatic regime for the mercury co-magnetometer. 

The mercury false EDM value given by Eq.\ \eqref{HgFalseEDM} is in practice times larger than the $d_{\rm Hg}$ experimental upper bound from direct searches for the Hg atomic EDM, $d_{\rm Hg} < 7.4 \times 10^{-30} e \, {\rm cm}$ \cite{Graner2016}, where the presence of a ~0.5 bar buffer gas reduces the size of the motional false EDM to $d^{\rm false}_{\rm Hg} < 10^{-31} e \, {\rm cm}$ \cite{Swallows2013,Graner2016} in this experiment. 

We will now give expressions for the mercury-induced false EDM in the case of more general magnetic nonuniformities described by the harmonic polynomial expansion (\ref{HarmonicPolynomialExpansion}).
From Eqs. (\ref{HgFalseEDM1},\ref{HgFalseEDM2}) we find
\begin{equation}
\label{HgFalseEDM3}
d^{\rm false}_{n \leftarrow {\rm Hg}} =
- \frac{\hbar \abs{\gamma_n \gamma_{\rm Hg}}}{2 c^2}
\sum_{l,m} G_{l,m} \langle \rho \Pi_{\rho, l, m} \rangle,
\end{equation}
where $\rho, z, \phi$ are the cylindrical coordinates and
\begin{equation}
\Pi_{\rho, l, m} = \cos \phi \, \Pi_{x, l, m} + \sin \phi \, \Pi_{y, l, m} = \partial_\rho \Sigma_{l+1, m}
\end{equation}
is the radial component of the mode $l,m$.
In table \ref{radialModes} we give expressions for the radial components of the first $m=0$ modes (see appendix \ref{appendixCylindrical} for more information on the harmonic polynomials in cylindrical coordinates). 

\begin{table}
\caption{
Radial components of the $l, m=0$ modes.
\label{radialModes}}
\begin{tabular}{ll}
$l$ & $\Pi_{\rho, l, m=0}(\rho, z)$ \\
\hline \hline
0  & $0$ \\
1  & $-\frac{1}{2} \rho$ \\
2  & $- \rho z$ \\
3  & $\frac{3}{8} \rho^3 - \frac{3}{2} \rho z^2$ \\
4  & $\frac{3}{2}\rho^3 z - 2 \rho z^3$ \\
5  & $-\frac{5}{16} \rho^5 + \frac{15}{4} \rho^3 z^2 -\frac{5}{2} \rho z^4 $ \\
6  & $-\frac{15}{8} \rho^5 z + \frac{15}{2} \rho^3 z^3 - 3 \rho z^5$ \\
7  & $\frac{35}{128} \rho^7 - \frac{105}{16} \rho^5 z^2 + \frac{105}{8} \rho^3 z^4 - \frac{7}{2} \rho z^6$
\end{tabular}
\end{table}

Next, we specify the formula (\ref{HgFalseEDM3}) in the case of a cylindrical chamber of radius $R$ and height $H$.
The origin of the coordinate system is at the center of the cylinder.
All $m \neq 0$ modes satisfy $\langle \rho \Pi_{\rho, l, m} \rangle = 0$ due to the average over $\phi$.
All even $l$ modes satisfy $\langle \rho \Pi_{\rho, l, 0} \rangle = 0$ due to the average over $z$.
Therefore, only the modes $\Pi_{\rho, l, 0}$ with $l$ odd contribute to the mercury-induced false EDM:
\begin{eqnarray}
\label{HgFalseEDM4}
d^{\rm false}_{n \leftarrow {\rm Hg}} & = &
- \frac{\hbar \abs{\gamma_n \gamma_{\rm Hg}}}{2 c^2}
\sum_{l \, {\rm odd}} G_{l,0} \langle \rho \Pi_{\rho, l, 0} \rangle \\
& = & \frac{\hbar \abs{\gamma_n \gamma_{\rm Hg}}}{8 c^2} R^2 \left[ G_{1,0} - G_{3,0} \left(\frac{R^2}{2}-\frac{H^2}{4} \right) \right. \\
\nonumber
& & \left. + G_{5,0} \left(\frac{5R^4}{16} - \frac{5R^2 H^2}{12} + \frac{H^4}{16} \right) + \cdots \right].
\end{eqnarray}

The motional false EDM of mercury induced by the linear gradient $G_{1,0}$ has been experimentally confirmed in \cite{Afach2015}, by applying an artificially large gradient. 
More recently we have also verified the effect induced by the cubic term $G_{3,0}$ with a dedicated measurement as reported in Sec. \ref{ExpCubic}. 
The motional false EDM is a dominant systematic effect that must be compensated for, and in order to determine the true EDM from experimental values one must extrapolate the measured EDM to zero gradient. 
An effective strategy for that extrapolation, used in the previous measurement \cite{Pendlebury2015}, takes advantage of neutron frequency shifts which are also sensitive to the gradients. 
We will review these frequency shifts in section \ref{FreqShifts} and explain the correction strategy using the gravitational shift in section \ref{CorrectionStrategy}. 

\subsection{Experimental verification of the false EDM induced by the cubic mode}
\label{ExpCubic}

In order to verify the accuracy of the predicted false EDM $d^{\rm false}$, a dedicated measurement was performed in the neutron EDM experiment at PSI using different magnetic field gradients. 
In a previous work \cite{Afach2015} we verified that a linear gradient $G_{1,0}$ produces a motional false EDM on the mercury as predicted by the theory. 
Here we extend this verification to the false EDM produced by the cubic mode $G_{3,0}$. 

In this measurement no neutrons were used, and the $^{199}$Hg precession frequency $f_{\rm Hg}$ was monitored while the applied electric field was periodically reversed: $E = \pm 120 \, {\rm kV} / 12 \, {\rm cm}$.
The measurements were performed in a series of standard cycles for which the sequence begins with the filling of the precession chamber with  spin-polarized Hg atoms.
The Hg spin is then flipped to a transverse direction (with respect to $B_0$) using a $\pi/2$ magnetic resonance pulse of 2~s duration.
A weak circularly polarized light beam is used to monitor the precessing transverse Hg spins by measuring the light power transmitted though the Hg medium.
Due to the spin-dependent part of the absorption coefficient, the transmitted power is modulated synchronously with the spin precession.
After recording the free-spin precession for 72 s, the cycle ends with the emptying of the precession chamber.
Cycles were repeated every 100 s, and the E-field was reversed in a $+--+$ pattern where every entry in the pattern consists of ten cycles.

The change in Hg precession frequency $\Delta f_{\rm Hg}$ correlated with the change in electric field $\Delta E$ was analyzed by averaging over many electric-field changes.
The pattern $+--+$ suppresses the effect of linear drifts in the Hg precession frequency due to slow changes of the magnetic field in the apparatus.
Periods during which the magnetic field changed rapidly (e.g.\ because of ramping superconducting magnets in neighboring experiments) were cut from the data analysis.

\begin{figure}
\includegraphics[width = \columnwidth]{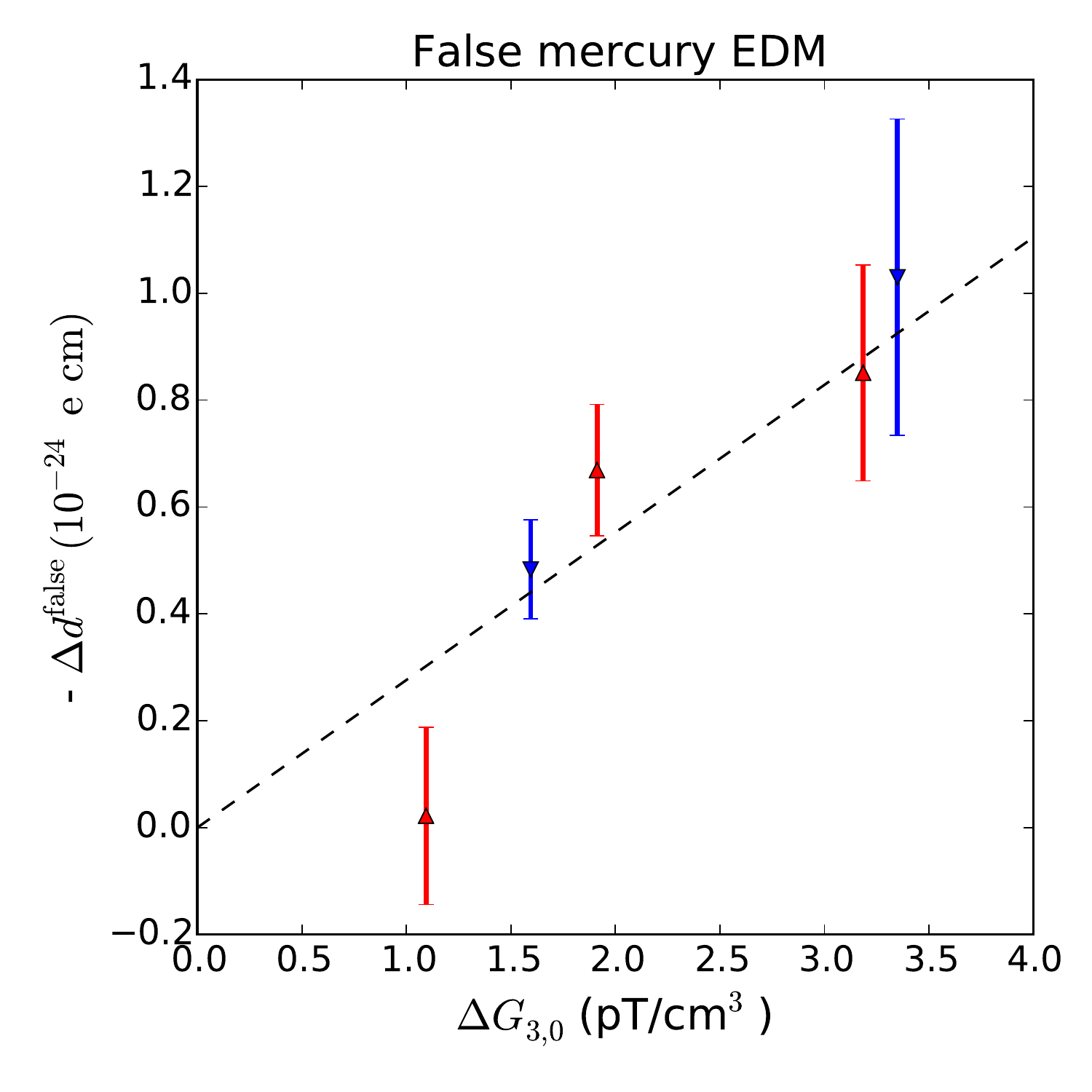}
\caption{\label{fig:CubicShiftExp}
Experimental verification of motional false EDM of mercury induced by a change of the cubic gradient $G_{3,0}$.
The frequency shift correlated with electric-field reversals was measured at $\pm 120$~kV. 
Red triangles pointing upwards (blue downwards) correspond to runs for which the $B_0$ field points upwards (downwards). 
The dashed line corresponds to the theoretical expectation given by Eq.\ \eqref{TheoCubic}. 
}
\end{figure}

We took data in a number  of different magnetic field configurations. 
To change the cubic mode $G_{3,0}$ we applied appropriate currents in trim-coils mounted around the precession volume. 
For each magnetic field configuration we calculate the false EDM as
\begin{equation}
d^{\rm false} = \frac{\pi \hbar}{2 |E|} \left( f_{{\rm Hg}, \uparrow \uparrow} - f_{{\rm Hg}, \uparrow \downarrow} \right) . 
\end{equation}
We selected pairs of runs that only differ by the value of the cubic mode. 
We report in Fig.\ \ref{fig:CubicShiftExp} the difference $\Delta d^{\rm false}$ between each pair as a function of the cubic mode difference $\Delta G_{3,0}$. 
The value $\Delta G_{3,0}$ is inferred by analyzing field maps. 
We plan to describe the field mapping device and the analysis of the recorded maps in a later publication. 

Figure \ref{fig:CubicShiftExp} also shows the theoretical expectation
\begin{equation}
\label{TheoCubic}
\Delta d^{\rm false} = - \frac{\hbar \gamma_{\rm Hg}^2}{8 c^2} R^2 \left(\frac{R^2}{2}-\frac{H^2}{4} \right) \Delta G_{3,0}. 
\end{equation}
The measurement is in good agreement with the theory. 
More details about this measurement can be found in the PhD thesis of S. Komposch \cite{Komposch2017}.


\subsection{Electric-field-independent frequency shifts}
\label{FreqShifts}

We will now discuss the frequency shifts unrelated to the electric field in situations where  the Larmor frequencies of the neutrons $f_{n}$ and mercury atoms $f_{\rm Hg}$ are measured in a weak magnetic field $B_0 = 1~\muT$. 

There are several known effects that could significantly shift the ratio $\mathcal{R} = f_{n}/f_{\rm Hg}$ from its unperturbed value $\abs{\gamma_{n} / \gamma_{\rm Hg}}$. 
For the purpose of the present discussion we write the combination of these effects as 
\begin{equation}
\label{Rshift}
\mathcal{R} = \frac{f_{n}}{f_{\rm Hg}} = \abs{\frac{\gamma_{n}}{\gamma_{\rm Hg}}}
\left( 1 + \delta_{\rm Grav} + \delta_{\rm T} + \delta_{\rm other} \right). 
\end{equation}
The term $\delta_{\rm Grav}$ is called the gravitational shift and $\delta_{\rm T}$ is the shift due to transverse magnetic fields. 
The last term, $\delta_{\rm other}$, accounts for shifts unrelated to the field uniformity. 
It includes the effect of Earth rotation \cite{Lamoreaux2007}, Ramsey-Bloch-Siegert shifts due to imperfect $\pi/2$ pulses, and light shifts induced by the UV light probing the mercury precession. 
A discussion of these effects, which in practice are sub-dominant, is beyond the scope of this article; they were briefly discussed in \cite{Afach2014}. 
The first two terms $\delta_{\rm Grav}$ and $\delta_{\rm T}$ are of interest here because they are induced by the magnetic-field nonuniformity. 

The gravitational shift $\delta_{\rm Grav}$ is the dominating shift in Eq.\ \eqref{Rshift}. 
As we already have mentioned when discussing gravitational depolarization, ultracold neutrons ``sag'' towards the bottom of the chamber quite significantly due to gravity. 
In contrast, the mercury atoms form a gas at room temperature that fills the precession chamber uniformly. 
This results in slightly different average magnetic fields for the neutrons and the atoms in the presence of a vertical field gradient.  
In the framework of the harmonic expansion of the field, the volume average of the vertical component is
\begin{equation}
\langle B_z \rangle = \sum_{l, m} G_{l,m} \langle \Pi_{z, l, m} \rangle. 
\end{equation}
For a cylindrical chamber all the terms with $m \neq 0$ vanish. 
Limiting the expansion to $l=3$, we have
\begin{eqnarray}
\nonumber
\langle B_z \rangle & & = G_{0,0} + G_{1,0} \langle z \rangle  \\ 
& & + G_{2,0} \langle -\rho^2/2 + z^2 \rangle 
+G_{3,0} \langle z^3 - \frac32 \rho^2 z \rangle. 
\end{eqnarray}
For both mercury atoms and neutrons we have
\begin{equation}
\langle \rho^2 \rangle = \frac{R^2}{2}. 
\end{equation}
For the mercury atoms we have
\begin{eqnarray}
& \langle z \rangle_{\rm Hg} \ & = 0, \\
& \langle z^2 \rangle_{\rm Hg} \ & = \frac{H^2}{12} \\
& \langle z^3 \rangle_{\rm Hg} \ & = 0. 
\end{eqnarray}
Therefore, the averaged field, which we call the $B_0$ field, is 
\begin{equation}
\label{B0Hg}
B_0 := \langle B_z \rangle_{\rm Hg} = G_{0,0} + G_{2,0} \left(\frac{H^2}{12} - \frac{R^2}{4} \right). 
\end{equation}
Now, for neutrons, the main difference when compared to atoms is that 
the center of mass $\langle z \rangle_{n}$ -- which we denote simply as $\langle z \rangle$ -- is significantly nonzero and negative. 
To calculate the ensemble average of higher powers of $z$, we approximate the neutron density $n(z)$ to be a linear function of $z$. We find 
\begin{eqnarray}
\label{z2z3}
\langle z^2 \rangle_{n} & \approx & \frac{H^2}{12}, \\
\langle z^3 \rangle_{n} & \approx & \frac{3 H^2}{20} \langle z \rangle. 
\end{eqnarray}
In reality the neutron density is not precisely a linear function of $z$. 
However, these expressions have been numerically verified to be accurate to better than a few percent for typical UCN spectra in storage vessels similar to
those used. 
Therefore, the expression of the field averaged by the neutrons is
\begin{eqnarray}
\nonumber
\langle B_z \rangle_n = & & G_{0,0} + G_{1,0} \langle z \rangle + G_{2,0} \left(\frac{H^2}{12} - \frac{R^2}{4} \right) \\
\label{B0n}
& & + G_{3,0} \left( \frac{3H^2}{20} - \frac{3R^2}{4} \right) \langle z \rangle. 
\end{eqnarray}

From Eq. \eqref{B0Hg} and \eqref{B0n} we deduce the gravitational shift 
\begin{equation}
\delta_{\rm Grav} = \frac{\langle B_z \rangle_{n}}{\langle B_z \rangle_{\rm Hg}} - 1 = \pm \frac{G_{\rm grav} \langle z \rangle}{|B_0|}, 
\end{equation}
where the $\pm$ sign refers to the direction of the magnetic field and 
the term $G_{\rm grav}$ is given by the following combination: 
\begin{equation}
G_{\rm grav} = G_{1,0} + G_{3,0} \left(\frac{3H^2}{20} - \frac{3R^2}{4} \right). 
\end{equation}

The second shift in Eq.\ \eqref{Rshift}, $\delta_{\rm T}$, arises from residual transverse field components $B_{\rm T}$. 
As mentioned above, the neutrons fall into the adiabatic regime of slow particles in a high field, and therefore the spins precess at a rate given by the volume average of the modulus of the field: 
\begin{equation}
f_n = \frac{\abs{\gamma_{n}}}{2 \pi} \langle | B | \rangle_{n} \approx \frac{\abs{\gamma_{n}}}{2 \pi} \left( |\langle B_z \rangle_{n}| + \frac{ \langle B_{\rm T}^2 \rangle}{2 |B_0|} \right). 
\end{equation}
The mercury atoms on the other hand fall into the nonadiabatic regime of fast particles in a low field, as a result of which the spins precess at a rate given by the vectorial volume average of the field: 
\begin{equation}
f_{\rm Hg} = \frac{\gamma_{\rm Hg}}{2 \pi} |\langle \vec{B} \rangle_{\rm Hg}| = \frac{\gamma_{\rm Hg}}{2 \pi} |B_0|. 
\end{equation}
Due to the fact that $\langle B_z \rangle_{n} \neq B_0$ is already accounted for by the gravitational shift, the expression for the transverse shift is simply 
\begin{equation}
\delta_{\rm T} = \frac{\langle B_{\rm T}^2 \rangle}{2 B_0^2}. 
\end{equation}
The expression for $\langle B_{\rm T}^2 \rangle$ in terms of the coefficients $G_{l,m}$ 
is given in appendix \ref{appendixTransverse}.

\subsection{Experimental verification of the gravitational and transverse shifts}
\label{ExpGravTransverse}

\begin{figure}
\includegraphics[width = \columnwidth]{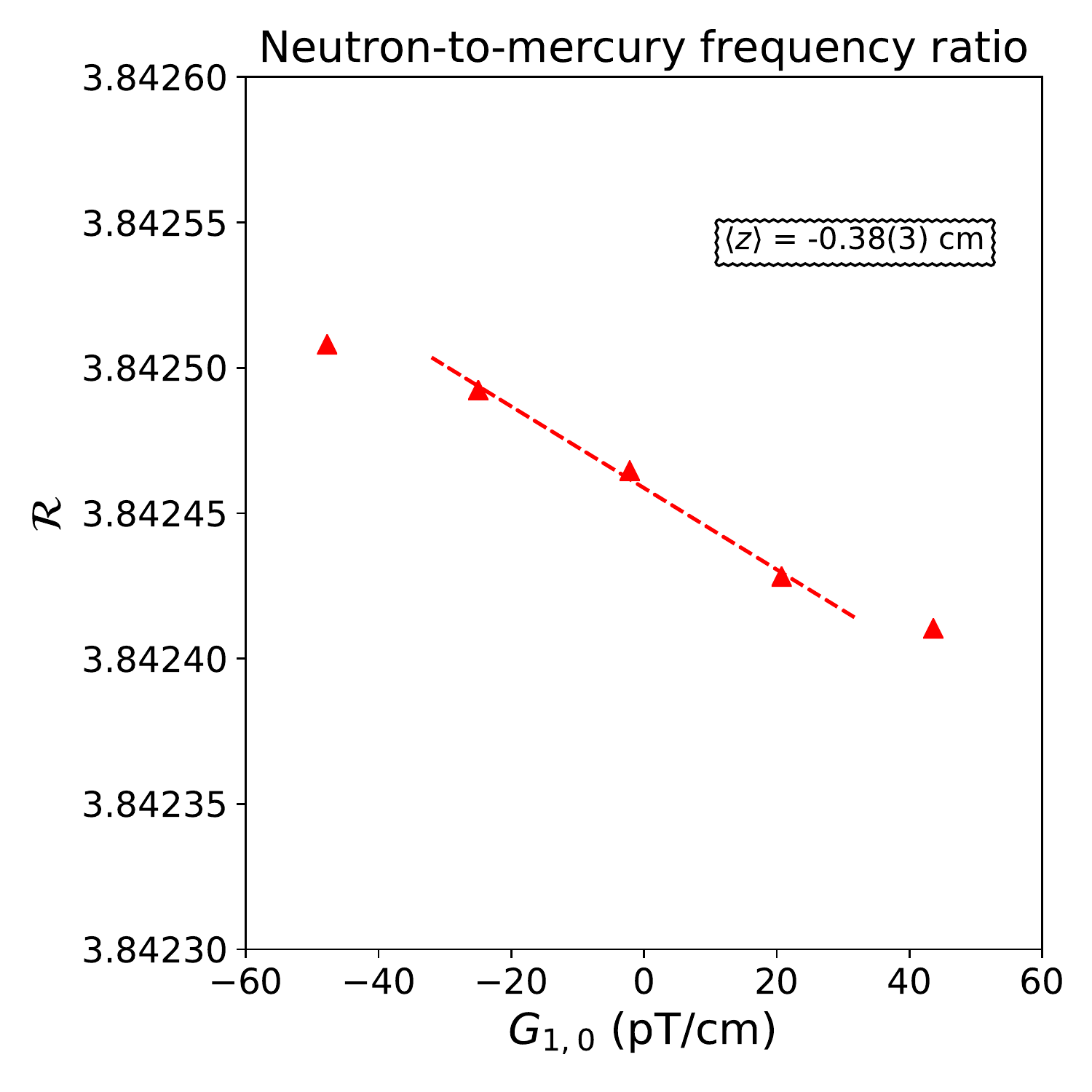}
\caption{\label{fig:Rcurve}
Experimental verification of the gravitational shift: neutron-to-mercury frequency ratio $\mathcal{R}$ as a function of an applied vertical gradient $G_{1,0}$. 
A linear fit to the data is performed (excluding the two points at large gradients) to extract $\langle z \rangle$. 
}
\end{figure}

In Fig.\ \ref{fig:Rcurve} we show a measurement of the ratio $\mathcal{R} = f_n/f_{\rm Hg}$ as a function of an applied vertical field gradient $G_{1,0}$. 
The underlying data are the same as those used to produce figure \ref{fig:G10scan}. 
We observe that the dependence of $\mathcal{R}$ versus the gradient is not quite linear. 
Fitting only the linear part we find $\langle z \rangle = -0.38(3)$~cm. 
The nonlinear behavior is primarily due to the phenomenon of {\em Ramsey wrapping} \cite{Harris2014, Afach2015_PRD}. Under the influence of gravity and in the presence of a vertical field gradient, the distribution of spin phases evolves in an asymmetric manner.  Ramsey's technique measures phase modulo $2\pi$, so a dominant tail on one side of the distribution can ``wrap around'' and effectively contribute to pulling the measured phase in the opposite direction to that which one would naively expect. (This effect is also very slightly enhanced by a subtle interplay between depolarization and frequency shift: the depolarization at large gradients acts differently upon the different energy groups, depolarizing the lowest-energy neutrons more quickly so that they contribute less to the frequency shift, thus effectively modifying $\langle z \rangle$; but the latter is a very minor addition.)  These complications, which are only relevant for large field gradients, have been neglected in the previous discussion.    

\begin{figure}
\includegraphics[width = \columnwidth]{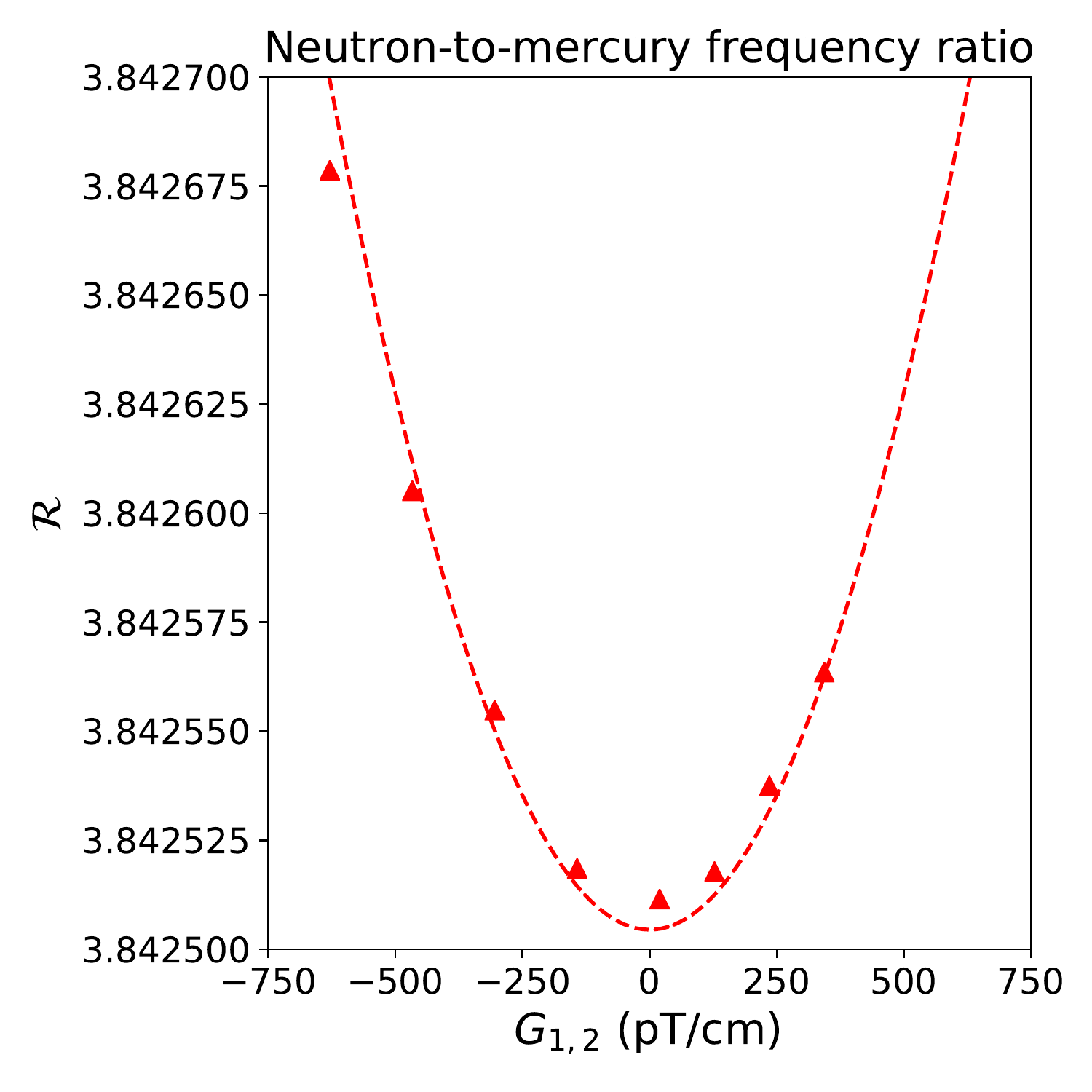}
\caption{\label{fig:Rcurve_G12}
Experimental verification of the transverse-gradient shift: neutron-to-mercury frequency ratio $\mathcal{R}$ as a function of applied transverse gradient $G_{1,2}$. 
The dashed line is a symmetric parabola with the constant term fitted to the data and the quadratic term fixed to the theoretical value. 
}
\end{figure}

Next we report on a dedicated experiment to verify the frequency shift due to a transverse field. 
The measurements were performed at PSI in October 2017. 
We varied the transverse field component using a combination of trim coils optimized to induce only the $G_{1,2}$ mode. 
Since the $G_{1,2}$ mode is purely transverse, the scalar Cs magnetometers could not be used to measure it; instead we used offline fluxgate maps of the trim coils to determine the value of $G_{1,2}$ as a function of the currents in the coils. 
Figure \ref{fig:Rcurve_G12} shows the $\mathcal{R}$ ratio as a function of $G_{1,2}$. 
We also carried out a similar test for the $G_{1,-2}$ mode, and that measurement is also in good agreement with the expected shift. 

\subsection{Correction strategy using the gravitational shift}
\label{CorrectionStrategy}

We now suggest a strategy to correct for the motional false EDM through use of the gravitational shift. 
We extend the method used in \cite{Pendlebury2015}, 
which neglected possible $l>1$ terms for the nonuniformity. 
Here we assume that the magnetic field can be described by the harmonic expansion up to $l=4$ and we neglect for the time being all terms $l>4$. 

For a given sequence of measurements with a fixed magnetic field configuration, the measured EDM is the sum of the true EDM and the false EDM, which can be written as:
\begin{equation}
d_n^{\rm meas} = d_n^{\rm true} + \frac{\hbar \abs{\gamma_{n} \gamma_{\rm Hg}}}{8 c^2} R^2 
\left[ G_{\rm grav} + G_{3,0} \left( \frac{R^2}{4} + \frac{H^2}{10} \right) \right]. 
\end{equation}
On the other hand, the $\mathcal{R}$ ratio measured for that magnetic field configuration is given by 
\begin{equation}
\mathcal{R} = \abs{\frac{\gamma_{n}}{\gamma_{\rm Hg}}} \left( 1 \pm \frac{G_{\rm grav} \langle z \rangle}{|B_0|} + \delta_{\rm T} + \delta_{\rm other} \right), 
\end{equation}
where the $+(-)$ sign refers to $B_0$ pointing upwards (downwards).
We define the corrected quantities $d_n^{\rm corr}$, $\mathcal{R}^{\rm corr}$ to be
\begin{equation}
d_n^{\rm corr} = d_n^{\rm meas} - \frac{\hbar \abs{\gamma_{n} \gamma_{\rm Hg}}}{8 c^2} 
R^2 \left( \frac{R^2}{4} + \frac{H^2}{10} \right) G_{3,0} 
\end{equation}
and
\begin{equation}
\mathcal{R}^{\rm corr} = \mathcal{R}/(1 + \delta_{\rm T} + \delta_{\rm other}).
\end{equation}
To calculate these, the magnetic-field related quantities $G_{3,0}$ and $\langle B_{\rm T}^2 \rangle$ are required. 
They can be measured offline by field mapping, if the reproducibility of the magnetic field configuration is sufficient. 

Then, we have 
\begin{equation}
d_n^{\rm corr} = d_n^{\rm true} + \frac{\hbar \abs{\gamma_{n} \gamma_{\rm Hg}}}{8 c^2} R^2 G_{\rm grav}
\end{equation}
and
\begin{equation}
\mathcal{R}^{\rm corr} = \abs{\frac{\gamma_{n}}{\gamma_{\rm Hg}}} \left( 1 \pm \frac{G_{\rm grav} \langle z \rangle}{|B_0|} \right).
\end{equation}
Therefore, 
\begin{equation}
d_n^{\rm corr} = d_n^{\rm true} + B_0 \frac{\hbar \gamma_{\rm Hg}^2}{8 c^2 \langle z \rangle} R^2 \left( \mathcal{R}^{\rm corr} - \abs{\frac{\gamma_{n}}{\gamma_{\rm Hg}}} \right). 
\end{equation}

Now, we have a set of ``points'' $(d_n^{\rm corr}, \mathcal{R}^{\rm corr})$, where each ``point'' corresponds to a different field configuration. 
It is important to get a set of points for both polarities of $B_0$. 
The so-called \emph{crossing-point analysis} simply consists of fitting these two series of points with two linear functions with opposite slope. It gives direct access to $d_n^{\rm true}$, since at the crossing point $d_n = d_n^{\rm true}$ and $\mathcal{R}^{\rm corr} = \abs{\frac{\gamma_{n}}{\gamma_{\rm Hg}}}$.  This technique was extended in \cite{Pendlebury2015} to include the nonlinearity arising from Ramsey wrapping, resulting in a far more satisfactory fit to the data.

Let us now make a few remarks. 
\begin{enumerate}
\item In principle, one could extract $G_{\rm grav}$ from offline field mapping or with real-time magnetometers around the precession chamber, and correct the false EDM on a point-by-point basis without using the crossing-point analysis. 
However, this requires an accuracy better than $1$~pT/cm for $G_{\rm grav}$ (corresponding to an error of $4.4 \times 10^{-27}$~e cm), which is beyond the reach of the current experimental setup. 
The accuracy of the determination of the gradients will be discussed quantitatively in the two aforementioned forthcoming articles. 
\item An experiment with a vertical stack of two chambers, rather than just one, could simply measure the gradient by taking the field difference between the top and bottom chambers. This would be an alternative to the gradient extracted via the gravitational shift. 
\item The crossing-point condition $\mathcal{R}^{\rm corr} = \abs{\frac{\gamma_n}{\gamma_{\rm Hg}}}$ allows an important cross-check of the analysis: $\mathcal{R}^{\rm corr}$ should agree with $\abs{\frac{\gamma_n}{\gamma_{\rm Hg}}}$ calculated from  independent measurements of $\gamma_n$ and $\gamma_{\rm Hg}$. 
\end{enumerate}

\subsection{The special case of a localized magnetic dipole}

The correction strategy presented in the previous paragraph compensates for the false EDM produced by a nonuniform field for all modes up to $l=4$. 
However, it does not perfectly compensate for the systematic effect generated by a localized magnetic dipole situated close to the precession chamber, as pointed out in \cite{Harris2006}. 
Indeed, the residual false EDM, after the correction procedure, is given by 
\begin{equation}
\label{dres_dipole}
d^{\rm res}_{n} = -\frac{\hbar \abs{\gamma_{n} \gamma_{\rm Hg}}}{2c^{2}} \left( \langle x B_{x}^{\rm dip} + y B_{y}^{\rm dip} \rangle + \frac{R^{2}}{4} \langle \frac{\partial B_z^{\rm dip}}{\partial z} \rangle \right), 
\end{equation}
where $(B_x^{\rm dip}, B_y^{\rm dip}, B_z^{\rm dip})$ is the magnetic field generated by the magnetic dipole. 
The first term corresponds to the systematic effect induced by the horizontal components of the dipole, and the second term arises from the correction procedure. 

When the dipole is situated on the axis below or above the cylindrical chamber, an analytical expression for Eq.\ \ref{dres_dipole} can be derived \cite{Pignol2012}. 
In general, however, for an arbitrary position of the magnetic dipole, Eq.\ \eqref{dres_dipole} has to be calculated numerically. 
Most critical are dipoles located on the circumference of the chamber. 

\begin{figure}
\includegraphics[width = \columnwidth]{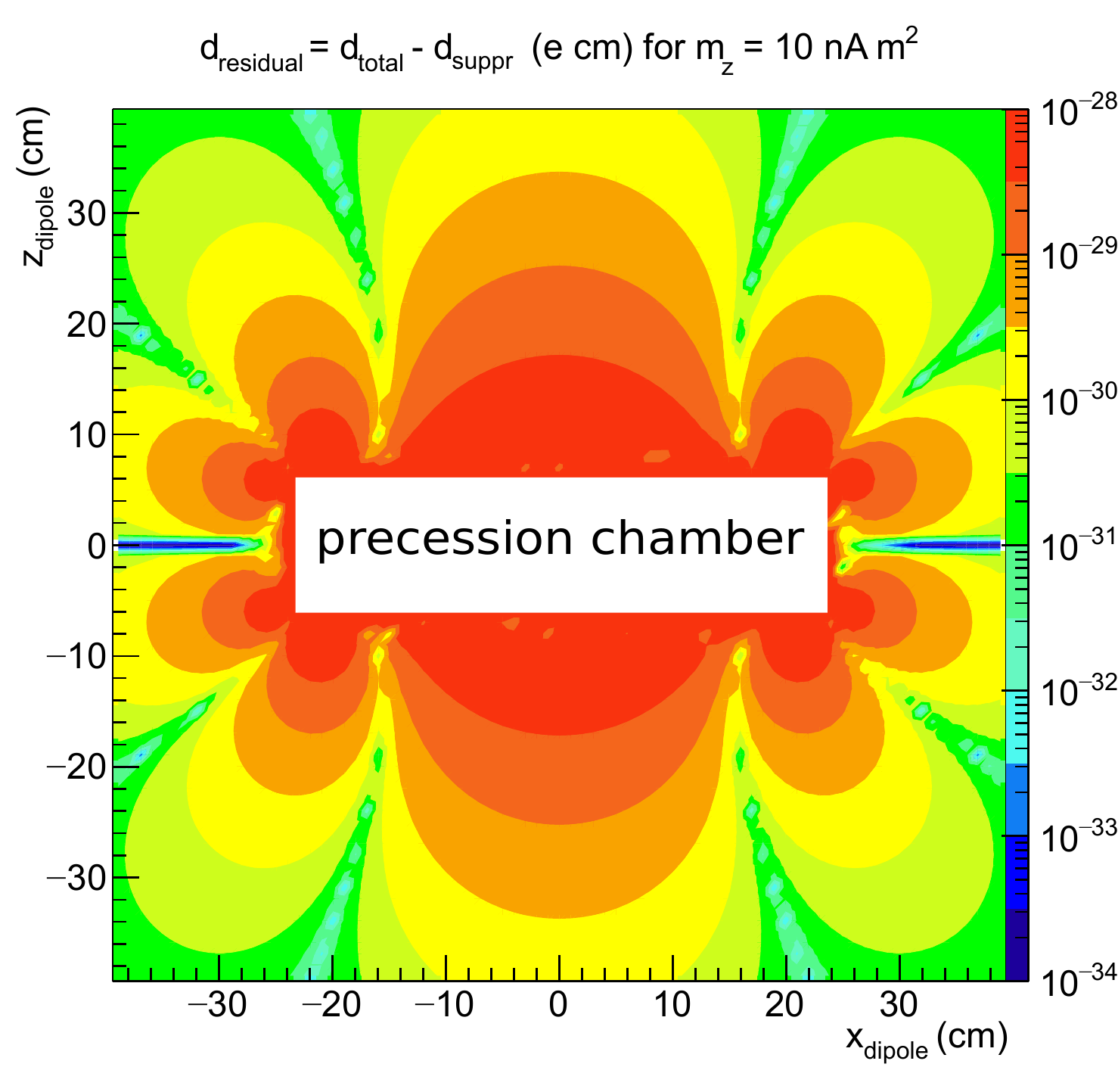}  
\caption{\label{fig:MagDipole}Absolute residual false EDM created by a dipole located in the vertical plane $y = 0$, with a magnetic moment aligned with $z$ and with $m_{z} = 10$~nA m$^{2}$, 
as a function of the position $(x,z)$ of the dipole. 
The white area corresponds to the volume of the chamber (diameter 47~cm and height 12~cm). 
}
\end{figure}

We show in Fig.\ \ref{fig:MagDipole} a numerical calculation of the false EDM generated by a dipole oriented along $z$, with a magnetic moment $m_{z} = 10$ nA m$^{2}$. 
This dipole corresponds to a speck of spherical iron dust with diameter $20 \ \upmu$m magnetized to saturation. 

\section{Summary and discussion}

In this paper we have discussed how magnetic field nonuniformities affect the statistical and systematic errors in the measurement of the neutron electric dipole moment. 

Concerning the statistical precision, the field uniformity must be sufficient to prevent the depolarization of ultracold neutrons during the precession time, which is as long as a few minutes. 
We have reviewed the main mechanisms of magnetic --~gravitational and intrinsic~-- depolarization. 
We have reported upon dedicated measurements of these effects, in particular using the UCN spin-echo technique to separate the intrinsic and gravitationally enhanced depolarization components. 

As far as systematic effects are concerned, we have focused the discussion on those related to the mercury co-magnetometer. 
In the previous literature, discussion about the false EDM effect in mercury was limited to linear gradients, although the case of localized dipoles was treated in \cite{Harris2006}, \cite{Clayton2011} and \cite{Pignol2012}. 
In this article we have extended the discussion to higher-order gradients. 
The theory for the motional false EDM is given in terms of a harmonic expansion. 
We have performed a dedicated measurement to verify the effect of the cubic mode in this expansion. 

We have in preparation two companion articles on the subject of magnetic field uniformity in the PSI nEDM experiment. 
The second episode of this trilogy will present the procedure to produce a uniform field \emph{in situ} with the help of an array of cesium magnetometers. 
The third article will present the offline characterization of the field uniformity through use of an automated mapping device. 

\begin{acknowledgements}
The experimental data were taken at PSI Villigen. 
We acknowledge the excellent support provided by the PSI technical groups and by various services of the collaborating universities and research laboratories. 
We  gratefully  acknowledge  financial  support  from
the  Swiss  National  Science  Foundation  through  projects
200020-137664  (PSI),  200021-117696  (PSI),  200020-
144473   (PSI),   200021-126562   (PSI),   200021-181996
(Bern),  200020-172639 (ETH) and 200020-140421 (Fri-
bourg);    and   from   STFC,   via   grants   ST/M003426/1,
ST/N504452/1 and ST/N000307/1. The LPC Caen and the
LPSC  Grenoble  acknowledge  the  support  of  the  French
Agence  Nationale  de  la  Recherche  (ANR)  under  reference ANR-09-BLAN-0046 and the ERC project 716651-NEDM.  The  Polish  collaborators  wish  to  acknowledge support from the National Science Center, Poland, under grant  no.   2015/18/M/ST2/00056.   P.  M.  acknowledges grant SERI-FCS 2015.0594.  This work was
also partly supported by the Fund for Scientific Research
Flanders  (FWO),  and  Project  GOA/2010/10  of  the  KU
Leuven.  In addition we are grateful for access granted to
the computing grid infrastructure PL-Grid.

\end{acknowledgements}

\appendix

\section{Harmonic polynomials in cylindrical coordinates}
\label{appendixCylindrical}
It is useful to derive the expressions of the harmonic modes in cylindrical coordinates 
$(\rho, \phi, z)$ with $x=\rho \cos \phi$ and $y=\rho \sin \phi$. 
The radial, azimuthal and vertical components respectively of the mode $l,m$ are given by 
\begin{eqnarray}
\Pi_{\rho, l, m} & = & \cos \phi \ \Pi_{x, l, m} + \sin \phi \ \Pi_{y, l, m} \\
 & = & \partial_\rho \Sigma_{l+1, m} \\
\Pi_{\phi, l, m} & = & - \sin \phi \ \Pi_{x, l, m} + \cos \phi \ \Pi_{y, l, m} \\
 & = & \frac{1}{\rho} \partial_\phi \Sigma_{l+1, m} \\
\Pi_{z, l, m} & = & \partial_z \Sigma_{l+1, m}. 
\end{eqnarray}

It is possible to write a simplified expression for the vertical component. 
Starting from Eq.\ \eqref{magneticPotential}, we have 
\begin{eqnarray}
& \Pi_{z, l, m}  = & C_{l+1,m}(\phi) \ \partial_z \left[ r^{l+1} P_{l+1}^m(c) \right] \\ 
\nonumber
& = & C_{l+1,m}(\phi) r^{l} 
\left[ (l+1) c P_{l+1}^m (c) + (1-c^2) \partial_c P_{l+1}^m(c) \right], 
\end{eqnarray}
where $c = \cos \theta$. 
Using the following known property of the associated Legendre polynomials,  
\begin{equation}
(c^2-1) \partial_c P_{l+1}^m(c) = (l+1) c P_{l+1}^m(c) - (l+1+m) P_{l}^m(c), 
\end{equation}
we arrive at
\begin{equation}
\Pi_{z, l, m} = C_{l+1,m}(\phi) (l+m+1) r^{l} P_{l}^m(\cos \theta). 
\end{equation}

It is also possible to write a simplified expression for the radial component, but only for the $m=0$ modes. 
In that case, 
\begin{eqnarray}
\Pi_{\rho, l, 0} & = & \frac{1}{l+1} \ \partial_\rho \left[ r^{l+1} P_{l+1}^0(c) \right] \\ 
\nonumber
& = & \frac{r^l}{l+1}  \sin \theta \left[ (l+1)P_{l+1}^0(c) - c \partial_c P_{l+1}^0(c)\right].  
\end{eqnarray}
We use the following property of the Legendre polynomials:
\begin{equation}
(l+1)P_{l+1}^0(c) - c \partial_c P_{l+1}^0(c) = - \partial_c P_{l}^0(c), 
\end{equation}
to find
\begin{equation}
\Pi_{\rho, l, 0} = \frac{r^l}{l+1} \frac{d}{d\theta} P_{l}^0 (\cos \theta). 
\end{equation}

An explicit calculation of the modes in cylindrical coordinates up to $l=3$ is shown in table \ref{adequateCylindrical}. 

\begin{table*}
\caption{
The basis of harmonic polynomials sorted by degree in cylindrical coordinates. 
\label{adequateCylindrical}}
\begin{tabular}{cc|ccc}
$l$ & $m$ & $\Pi_\rho$ & $\Pi_\phi$  & $\Pi_z$ \\
\hline \hline
$0$ & $-1$  & $\sin \phi$         & $\cos \phi$         & $0$          \\
$0$ & $0$  & $0$                 & $0$                 & $1$          \\
$0$ & $1$  & $\cos \phi$         & $-\sin \phi$        & $0$          \\
\hline
$1$ & $-2$  & $\rho \sin 2\phi$   & $\rho \cos 2\phi$   & $0$               \\
$1$ & $-1$  & $z \sin \phi$       & $z \cos \phi$       & $\rho \sin \phi$  \\
$1$ & $0$  & $-\frac12 \rho$     & $0$                 & $z$               \\
$1$ & $1$  & $z \cos \phi$       & $-z \sin \phi$      & $\rho \cos \phi$  \\
$1$ & $2$  & $\rho \cos 2\phi$   & $- \rho \sin 2\phi$ & $0$               \\
\hline
$2$ & $-3$ & $\rho^2 \sin 3\phi$  & $\rho^2 \cos 3\phi$ & $0$          \\
$2$ & $-2$ & $2 \rho z \sin 2\phi$     & $2\rho z \cos 2\phi$        & $\rho^2 \sin 2\phi$   \\
$2$ & $-1$ & $\frac14 (4z^2-3\rho^2)\sin \phi$ & $\frac14 (4z^2-\rho^2)\cos \phi$ & $2\rho z \sin \phi$ \\
$2$ & $0$ & $- \rho z$           & $0$                 & $-\frac12 \rho^2 + z^2$      \\
$2$ & $1$ & $\frac14 (4z^2-3\rho^2)\cos \phi$ & $\frac14 (\rho^2-4z^2)\sin \phi$ & $2 \rho z \cos \phi$ \\
$2$ & $2$ & $2 \rho z \cos 2\phi$    & $-2 \rho z\sin 2\phi$     & $\rho^2 \cos 2\phi$    \\
$2$ & $3$ & $\rho^2 \cos 3\phi$ & $-\rho^2 \sin 3\phi$ & $0$         \\
\hline
$3$ & $-4$ & $\rho^3 \sin 4\phi$      & $\rho^3 \cos 4\phi$        & $0$            \\
$3$ & $-3$ & $3 \rho^2 z \sin 3\phi$ & $3 \rho^2 z \cos 3\phi$   & $\rho^3 \sin 3\phi$    \\
$3$ & $-2$ & $\rho (3z^2-\rho^2) \sin 2\phi$ & $\frac12 \rho (6z^2-\rho^2) \cos 2\phi$ & $3 \rho^2 z \sin 2\phi$     \\
$3$ & $-1$ & $\frac 14z(4z^2-9\rho^2)\sin \phi$ & $\frac 14z(4z^2-3\rho^2) \cos \phi$   & $\rho(3 z^2- \frac 34 \rho^2) \sin \phi$    \\
$3$ & $0$ & $\frac38\rho(\rho^2-4z^2)$  & $0$  & $\frac12z(2z^2-3\rho^2)$  \\
$3$ & $1$ & $\frac 14z(4z^2-9\rho^2)\cos \phi$ & $\frac 14z(3\rho^2-4z^2)\sin \phi$ & $\rho(3z^2 - \frac 34 \rho^2)\cos \phi$ \\
$3$ & $2$ & $\rho(3z^2-\rho^2)\cos 2\phi$    & $\frac12\rho(\rho^2-6z^2) \sin 2\phi$   & $3 \rho^2 z \cos 2\phi$  \\
$3$ & $3$ & $3 \rho^2 z \cos 3\phi$  & $-3 \rho^2 z \sin 3\phi$   & $\rho^3 \cos 3\phi$   \\
$3$ & $4$ & $\rho^3 \cos 4\phi$      & $- \rho^3 \sin 4\phi$      & $0$            \\
\hline
\end{tabular}
\end{table*}

\section{Transverse field components}
\label{appendixTransverse}

In this appendix we give the expression for the mean squared transverse field, 
\begin{equation}
\av{B_{\rm T}^2} = \av{(B_x-\av{B_x})^2+(B_y-\av{B_y})^2}, 
\end{equation}
in terms of the generalized gradients $G_{l,m}$ up to order $l=3$ for a cylindrical precession chamber of radius $R$ and height $H$. 

It can be expressed as a sum of four contributions: 
\begin{equation}
\av{B_{\rm T}^2} = \av{B_{\rm T}^2}_{\rm LO} + \av{B_{\rm T}^2}_{\rm 2O} + \av{B_{\rm T}^2}_{\rm 3O} + \av{B_T^2}_{\rm 3I1}. 
\end{equation}
The linear-order contribution is 
\begin{eqnarray}
\nonumber
\av{B_{\rm T}^2}_{\rm LO} = & & \frac{R^2}{2} (G_{1,-2}^2+ G_{1,2}^2+\frac14 G_{1,0}^2) \\
& & + \frac{H^2}{12} (G_{1,-1}^2+G_{1,1}^2).
\end{eqnarray}
The quadratic-order contribution is 
\begin{eqnarray}
\nonumber
\av{B_{\rm T}^2}_{\rm 2O} 
& = & \frac{R^4}{3} (G_{2,-3}^2+G_{2,3}^2) \\ 
\nonumber
& & + \frac{R^2 H^2}{12}(2G_{2,-2}^2 + 2G_{2,2}^2 + \frac12 G_{2,0}^2) \\
& & + \left( \frac{R^4}{24}+\frac{H^4}{180} \right) (G_{2,-1}^2+G_{2,1}^2).
\end{eqnarray}
The cubic-order contribution is 
\begin{eqnarray}
\nonumber
\av{B_{\rm T}^2}_{\rm 3O} 
& = & \frac{R^6}{4} (G_{3,-4}^2+G_{3,4}^2) \\ 
\nonumber
& & + \frac{R^4 H^2}{4} (G_{3,-3}^2+G_{3,3}^2) \\ 
\nonumber
& & + \left( \frac{5R^6}{32} - \frac{R^4H^2}{8} + \frac{9R^2 H^4}{160} \right) (G_{3,-2}^2+G_{3,2}^2) \\
\nonumber
& & + \left( \frac{5R^4H^2}{64} - \frac{3R^2H^4}{160} + \frac{H^6}{448} \right) (G_{3,-1}^2+G_{3,1}^2) \\
& & + \left( \frac{9R^6}{256} - \frac{R^4H^2}{32} + \frac{9 R^2H^4}{640} \right) G_{3,0}^2.
\end{eqnarray}
Finally, there is the interference term between the linear and cubic modes: 
\begin{eqnarray}
\av{B_{\rm T}^2}_{\rm 3I1} 
\nonumber
& = & \left(- \frac{R^4}{2}+\frac{R^2H^2}{4} \right) ( G_{1,-2} G_{3,-2} + G_{1,2} G_{3,2} + \frac14 G_{1,0} G_{3,0} ) \\
& & + \left( -\frac{R^2H^2}{8} + \frac{H^4}{40} \right) (G_{1,-1} G_{3,-1} + G_{1,1} G_{3,1}).
\end{eqnarray}
Note that the quadratic modes do not interfere with the linear and cubic modes.

\end{document}